\documentclass[prd,aps,floats,eqsecnum,nofootinbib]{revtex4}
\usepackage{amsmath,amssymb,verbatim,epsfig,psfig,graphicx,rotating}
\newcommand{\be}{\begin{equation}}
\newcommand{\ee}{\end{equation}}
\newcommand{\bea}{\begin{eqnarray}}
\newcommand{\eea}{\end{eqnarray}}
\begin{document}
\title{\bf
THE SELF-GRAVITATING GAS IN THE PRESENCE OF DARK ENERGY}
\author{\bf  H. J. de Vega, J. A. Siebert \\ 
Laboratoire de Physique Th\'eorique et Hautes Energies, \\
Universit\'e Paris VI et Paris VII, Tour 16, 1er \'etage, \\ 4, Place Jussieu
75252 Paris cedex 05, France. \\
Laboratoire Associ\'e au CNRS UMR 7589.}
\begin{abstract}
The  non-relativistic self-gravitating gas in thermal equilibrium in
the presence of a positive cosmological constant  $ \Lambda $  (dark energy) is
investigated. The dark energy introduces a force  pushing outward all
particles with strength proportional to their distance to the center
of mass. We consider the statistical mechanics of the self-gravitating
gas of $N$ particles in a volume $V$ at thermal equilibrium in the
presence of $ \Lambda $. It is shown that the thermodynamic limit
exists and is described by the mean field equations provided $N, \; V
\to \infty $ with $N/V^{\frac13} $ fixed {\bf and} $ \Lambda \, V^{\frac23} $
fixed. That is,  $ \Lambda \to 0 $ for $N, \; V\to \infty $. The
case of $\Lambda$ fixed and $N, \; V\to \infty $ is solved too.
We solve numerically the mean field equation for
spherical symmetry obtaining an isothermal sphere for $\Lambda>0$.
 The  particle distribution turns to flatten compared with 
the  $\Lambda = 0 $ case. Moreover, the particle density increases
with the distance  when the cosmological constant dominates. There
is a bordering case with uniform density. The density contrast between
the center and the boundary may be significatively reduced by the dark
energy. In addition, the critical point associated to the collapse
(Jeans') phase 
transition is pushed towards higher values of $ N/[T \; V^{\frac13}] $
 by the presence of $\Lambda > 0 $. The nature and the behaviour near
 the critical points is not affected by the presence of $\Lambda > 0 $. 
\end{abstract} 
\date{\today} 
\maketitle
\tableofcontents

\section{Introduction}

The self-gravitating gas in thermal equilibrium has been thoroughly
studied since many years\cite{hidro,gasn,gas2,gas3,otros}. As a
consequence of the long range attractive Newton force, the
selfgravitating gas admits a consistent thermodynamic limit 
  $ N, V \to   \infty $ with $ \frac{N}{V^{\frac13}} $ fixed. In this limit,
  extensive thermodynamic quantities like energy, free energy, entropy
  are proportional to $N$\cite{hidro,gasn,gas2,gas3,otros}.

We investigate in this paper how the presence of a cosmological constant 
affects the properties of the nonrelativistic self-gravitating gas
in thermal equilibrium.

The cosmological constant modifies the  
Poisson equation adding a negative term in the r. h. s. This
introduces an extra  force {\bf pushing outwards} all particles with a
strength proportional to their distance to the center of mass. This outward
force is the nonrelativistic version of the exponentially fast
expansion in the de Sitter universe. It is an antigravitational effect.

We consider the statistical mechanics of the
self-gravitating gas in the presence of a cosmological constant and
show that it is described by the mean field equation in the
thermodynamic limit. We derive the mean field equation in the
thermodynamic limit of large 
number $N$ of particles and large volume $V$ with $N/V^{1/3}$ and $ \Lambda \,
V^{\frac23} $ fixed. Furthermore, we consider the case $ \Lambda $
fixed for $N$ and $V$ going to infinity with $N/V^{1/3}$  fixed.

As known, the usual thermodynamic limit $N, \; V \to \infty $ with
$N/V$ fixed leads to collapse in absence of the cosmological constant,
while the dilute limit keeping $N/V^{1/3}$ fixed leads to a gaseous
phase \cite{gasn,gas2,gas3}. Here, we consider a {\bf vanishing} $
\Lambda $ for $ V \to \infty $ keeping $ \Lambda \, V^{\frac23} $
fixed as well as  $N/V^{1/3}$.

The meaning of the condition $ \Lambda \, V^{\frac23} $ for $ N, \; V
\to \infty $ goes as follows. The dark energy in a volumen $V$ is of
the order $ V \; \Lambda $ while the thermal (kinetic) energy (as well
as the interaction energy between the particles) is of the order $
N \; T $. Since $ N \sim V^{1/3},  \; \Lambda \, V^{\frac23} $ fixed
implies that the ratio of dark and thermal energies is fixed for $ N
\to \infty $. 

When $ \Lambda $ and $ N/V^{1/3}$ are kept fixed for  $N, \; V \to \infty
$ the cosmological constant dominates over self-gravity. The result is a set
of non-gravitating particles in a harmonic oscillator with negative
squared frequencies. This system is exactly solvable as we show.

It is convenient to define the dimensionless parameter:
\be\label{defxi}
\xi   \equiv 2 \, \Lambda\, G\, m \; \frac{L^2}{T}  \; . 
\ee
All physical magnitudes are expressed in terms of $\xi$ and 
$$ 
\eta= G\, m^2 \, N /[T \, L] \; ,
$$
the dimensionless parameter used in ref.\cite{gasn}.

We solve the mean field equations numerically for spherical symmetry. 
That is, we find the isothermal sphere in the presence of a non-zero
cosmological constant. 

We find that this  `antigravitational effect' of $\Lambda $ {\bf flattens} the
particle distributions in thermal equilibrium. The density contrast
(ratio of the density at the center and at the surface of the sphere)
{\bf decreases} for increasing $\Lambda $. The particle density at the
boundary is conformally translated as a function of $ \eta $ by the presence of
$\Lambda $ (see fig. \ref{ps}). 

In absence of cosmological constant the particle density is a
monotonically decreasing function of the radial distance $R$ \cite{hidro}
-\cite{gasn}. We find in the presence of a nonzero $\Lambda $ 
that the particle density may be either monotonically  decreasing {\bf
  or}  monotonically increasing with $R$. The particle density grows
with  $R$ when  $\Lambda $ dominates over the selfgravitational attraction. 

In the special case $ \eta= \xi $ the gravitational self-interaction
exactly compensates the cosmological constant. As a result the gas
becomes a perfect gas with uniform density.

The position of the collapse (Jeans') phase transition is pushed by
the cosmological constant to higher values of $ \eta $. Namely, lower
temperatures or volumes or higher number of particles. The nature of
the phase transition is not affected by the presence of $\Lambda > 0$
and  the behaviour of the physical magnitudes does not change far from
the phase transition (see figs. \ref{freeener}, \ref{poten} and
\ref{entropy}).  
Far from the collapse point, the energy and the free energy are basically 
shifted by the presence of $\Lambda > 0 $.

The thermodynamic quantities: free energy, energy and entropy turn to
be proportional to the number of particles in the thermodynamic
limit. We compute them as functions of $ \eta $ and $ \xi $. 
We also consider the limiting case  $ \xi \gg \eta$ which is solved
analytically in sec. V. This corresponds to keep $\Lambda$ fixed for $
V \to \infty $.  

As for $\Lambda=0$ the gas collapses in the canonical ensemble when
the compressibility diverges and becomes negative. This happens for a
value of  $ \eta $  smaller than the one where the specific heat is singular.

This article is organized as follows: in sec. 2 we recall the
equations of motion for nonrelativistic particles in the presence of a
cosmological constant. In section 3 we present the self-gravitating
gas in the presence of $\Lambda > 0 $; first in a hydrostatic approach,
 then the statistical mechanics and finally the mean field approach
 including the spherically symmetric case 
with  the calculation of the physical magnitudes (energy,
 free energy, entropy, surface density, pressure contrast, specific
 heat and compressibility).
 In sec. 4 we present the
 physical results for the the self-gravitating gas in the presence of
 $\Lambda > 0 $: phase structure and  particle density behaviour in the
 various cases. 
In sec. 5 we consider the exactly solvable case where the cosmological
constant dominates overwhemly over the self-gravitation. We present
our conclusions in sec. 6.

\section{Nonrelativistic gravity in the presence of the cosmological constant}

The energy-momentum tensor for massive particles plus dark energy (the
cosmological constant) takes the form
\begin{equation} \label{tmunu}
T^{\alpha}_{\beta} = \Lambda \; \delta^{\alpha}_{\beta} + \rho \;
\delta^{\alpha 0} \; \delta_{\beta 0} \; ,
\end {equation}
where $ \Lambda $ stands for the cosmological constant, $ \rho $ for
the energy density of the massive particles, 
$ 0 \leq \alpha, \beta \leq 3 $ and we assume nonrelativistic matter so
its pressure can be neglected compared with its rest mass. 
\noindent
The Einstein equations take the form
\begin{equation} \label{eins}
R^{\alpha}_{\beta} = 8\pi \, G \left(T^{\alpha}_{\beta} -
\frac{\delta^{\alpha}_{\beta}}2 \; T \right)\; ,
\end {equation}
where $ R^{\alpha}_{\beta} $ stands for the Ricci tensor and $ T
\equiv T^{\alpha}_{\alpha} $. We find from eq.(\ref{tmunu}) that
$$
T = 4 \, \Lambda +  \rho \quad \mbox{and}\quad T^0_0 - \frac12 \, T =
\frac12 \rho - \Lambda \; .
$$
We are interested in the non-relativistic limit where gravitational
fields are weak and we have for the metric \cite{llc}
$$
g_{00} = 1 + 2 \; V \quad , \quad g_{ik} = - \delta_{ik} \; ,
$$
where $V$ stands for the gravitational potential. One has 
for the $00$ component of the Ricci tensor\cite{llc}
$$
R_0^0 = \nabla^2 V \; .
$$
Therefore, the $00$ component of the  Einstein equations (\ref{eins})
becomes\cite{pee} 
\begin{equation} \label{poiL}
\nabla^2 V = 4\pi \, G \, \rho -  8\pi \, G \, \Lambda \; .
\end {equation}
For zero cosmological constant we recover the usual Poisson equation,
as it must be. 
Eq.(\ref{poiL}) can be written in an integral form,
\begin{equation} \label{potentieltot}
V({\vec q}) = V_{self}({\vec q})+V_{dark}({\vec q})
=-G \int \frac{\rho({\vec q'}) \; d^3 q'}{|{\vec q}-
{\vec q'}|}  - \frac{4\pi \, G \, \Lambda}{3} \, q^2 \; ,
\end{equation}
where $V_{self}$ is the contribution of the massive particles and
$V_{dark}$ is the contribution of the dark energy. Here we choose the
center of mass of the particles as origin of the coordinates. 
Linear terms in $ {\vec q} $ of the type
\begin{equation} \label{terl}
 {\vec a} {\cdot} {\vec q}
\end {equation}
with $\vec a $ an arbitrary  constant vector, are ignored in $ V({\vec q}) $
since they correspond to a 
constant external gravitational field unrelated to the cosmological
constant. [Notice that such terms have zero laplacian and are not
  determined by eq.(\ref{poiL})].
Such external field will accelerate the particles in the direction $
{\vec a} $.  Changing the origin of coordinates, terms linear in $ {\vec
  q} $ can be added to eq.(\ref{potentieltot}). 
Additional constant terms are clearly irrelevant. 

\bigskip

The mass density takes the following form for a distribution of point
particles at rest, 
$$
\rho({\vec q}) = \sum_i m_i \; \delta({\vec q}-{\vec q}_i) \; .
$$
Here, $ m_i $ stands for the mass of the particle at the point $ {\vec
q}_i $. 
The gravitational potential thus becomes,
$$
V({\vec q}) = -G \sum_i \frac{m_i}{|{\vec q}-{\vec q}_i|} -
\frac{4\pi \, G \, \Lambda}{3} \, q^2 \; . 
$$
and the gravitational field at the point $ \vec q $,
\begin{equation} \label{campog}
{\vec g } = - \nabla V({\vec q}) = - G \sum_i m_i \; \frac{ {\vec
q}-{\vec q}_i }{|{\vec q}-{\vec q}_i|^3} + \frac{8\pi \, G \,
\Lambda}{3} \; {\vec q}
\end {equation}
The potential energy of such a system of non-relativistic
self-gravitating particles in the presence of a cosmological constant
$ \Lambda $ thus takes the form,
\begin{equation} \label{enerG}
{\cal U} = -G \sum_{i<j} \frac{m_i \; m_j}{|{\vec q}_i-{\vec q}_j|}
- \frac{4\pi \, G \, \Lambda}{3} \,\sum_i m_i \; q_i^2 
\end {equation}
Therefore, the Hamiltonian can be written as
\begin{equation} \label{hamil}
H = \sum_i \frac{p_i^2}{2 \, m_i^2} + {\cal U}
\end {equation}
where $ p_i $ stands for the momentum of the $i$-th particle. 
The cosmological constant contribution to the potential energy
decreases for increasing values of the particle distances  $ r_i $ to
the center of mass. As stated above, in absence of external fields,
the terms like eq.(\ref{terl}) are absent in the potential
energy. Anyway, in  the center of mass frame such linear terms
identically vanish  upon summing over the particles in eq.(\ref{enerG}).

Therefore, the gravitational effect of the  cosmological
constant on particles amount to push them outwards. Equivalently,
the last term of the gravitational field eq.(\ref{campog}) points
{\bf outward}. It can be then said that the cosmological constant has an
{\bf antigravitational} effect. 

For simplicity we shall consider all particles with the same mass
$m$. The case of unequal masses may be worked out generalizing the
work in ref.\cite{gas3} to nonzero cosmological constant.

Dark energy introduces a kind of centrifugal force always pushing
out. Notice that the last term in eq.(\ref{enerG}) has the structure
of  a rotational energy but with negative sign. Also, it contains 
the distance squared of the particles to the origin and not the
distance squared to an axis as it is the case for rotational energy.

It is useful to consider the case of just one particle in the
presence of the cosmological constant in order to obtain physical insights.
We obtain from eqs.(\ref{enerG}) and (\ref{hamil}),
$$
{\ddot{\vec q}}=\frac{8\pi \, G \, \Lambda}{3} \, {\vec q} \; .
$$
This is an oscillator equation with imaginary frequency
and solution,
\begin{equation} \label{tray}
{\vec q}(t) = {\vec q}(0)\; \cosh Ht + \frac1{H} \; {\dot{\vec
q}}(0)\; \sinh Ht \; ,
\end {equation}
where $ H \equiv \sqrt{\frac{8\pi \, G \, \Lambda}{3}} $. 
The particle runs away exponentially fast in time. 

It is interesting to compare this non-relativistic results with the
full relativistic geometry in the presence of the cosmological
constant. The exact solution of the Einstein equations (the full
solution, not just the nonrelativistic limit) for the
energy-momentum eq.(\ref{tmunu}) with $ \rho = 0 $ is the de Sitter universe,
$$
ds^2 = dt^2 - e^{2 \, H\, t} \; (d {\vec x})^2 \; .
$$
Comparing the non-relativistic trajectories eq.(\ref{tray}) with the
exact relativistic geodesics in de Sitter space-time (see for example
\cite{plb}) one sees that both exhibit the same exponential runaway
behaviour. Therefore, the nonrelativistic approximation keeps the
essential features of the particle motion in de Sitter space-time.

\section{The self-gravitating gas with $ \Lambda > 0$}

We first consider the hydrostatic description of the self-gravitating gas in 
thermal  equilibrium in the presence of the cosmological
constant. Next, we present  the statistical mechanics of the
self-gravitating gas in the presence of the cosmological constant
and derive the mean field approach to it. The two
approaches turn to be equivalent when the number of particles tends to
infinity. 
  
\subsection{Hydrostatics of the self-gravitating gas with $ \Lambda
\neq 0$}

Let us consider the hydrostatic equilibrium condition \cite{llms} for
a self-gravitating gas in the presence of the cosmological constant.
\begin{equation} \label{equih}
\nabla P({\vec q}) = - \rho({\vec q}) \; \nabla V({\vec q})\; , 
\end {equation}
where $ P({\vec q}) $ stands for the pressure at the point $ \vec q
$. Let us assume the ideal gas equation of state  in local form (see
\cite{hidro,gasn,gas2}) 
\begin{equation} \label{gasi}
P({\vec q}) = \frac{T}{m} \; \rho({\vec q}) \; .
\end {equation}
Combining eqs.(\ref{equih}) and (\ref{gasi}) yields the Boltzmann law
for the particle density
\begin{equation} \label{Boltzdens}
 \rho({\vec q}) =  \rho_0 \; e^{ - \frac{m}{T} \, V({\vec q})}
\end {equation}
where $ \rho_0$ is a constant, $V$ being the volume
of the system. The normalization of density requires that
\begin{equation} \label{vincu}
\int_V  d^3{\vec q}  \; \rho({\vec q})  = mN \; .
\end {equation}
Inserting eq.(\ref{Boltzdens}) into the  Poisson equation
(\ref{poiL}) yields, 
\begin{equation} \label{ecgral}
\nabla^2 V = 4\pi \, G \, \left(
\rho_0 \; e^{ - \frac{m}{T} \, V({\vec q})} -  2 \, \Lambda \right) \; . 
\end {equation}
It is convenient to introduce dimensionless coordinates 
$ {\vec r} \equiv {\vec q}/L $ where $L\equiv V^{1/3} $ stands for the
linear size of the system. Eq.(\ref{ecgral}) takes thus the form,
\begin{equation} \label{ecsd}
\nabla^2 \Phi({\vec r}) + 4\pi \left(\eta \, e^{\Phi({\vec r})} - \xi
\right) =0 \; , 
\end {equation}
where $ \Phi({\vec r})  \equiv - \frac{m}{T} \, V({\vec q}) +
\log\frac{\rho_0 \; V}{m \, N} $, 
\begin{equation} \label{etaxi}
\eta= \frac{G\, m^2 \, N}{T \, L} \quad \mbox{and} \quad \xi   \equiv
2 \, \Lambda\, G\, m \, \frac{L^2}{T}  \; . 
\end {equation}
The mass density now becomes $\rho({\vec r}) =
e^{ \Phi({\vec r})} $ and the constraint (\ref{vincu}) takes the form, 
\begin{equation} \label{vincu2}
\int_{unit~volume}  d^3 {\vec r} \; \;  e^{ \Phi({\vec r})} = 1 \; .
\end {equation}
The limit for a large number of particles  $ N $ is well defined
provided we choose\cite{gasn}
$$
N \to \infty \quad , \quad L \to \infty \quad , \quad \Lambda \to 0 \;, 
$$
with 
$$
\frac{N}{L} = \mbox{fixed} \quad , \quad \Lambda \, L^2 = \mbox{fixed}
\; .
$$
It should be mention that an equation similar to eq.(\ref{ecsd}) appeared
in two space dimensions describing (multi)-vortices in the
Ginzburg-Landau or Higgs model in the limit between superconductivity
of type I and II \cite{vortex}. However, for the vortex case one has $
 \xi = \eta < 0 $. The negative sign is associated to the fact
that like charges repel each other in electrodynamics while masses
attract each other in gravity. 

In the limiting case $\xi=0$ eq.(\ref{ecsd}) becomes the well known
 Lane-Emden equation in the absence of cosmological
constant both in the hydrostatic\cite{hidro} and statistical mechanics
approaches \cite{gasn,gas2}.

Notice that $\Lambda$ plays the same role as the
subtraction of a constant density in the Jeans' swindle \cite{jeans}.

\subsection{The Partition function of the self-gravitating gas with $ \Lambda
\neq 0$}

We present now the statistical mechanics 
of the self-gravitating gas in the presence of the cosmological
constant in the canonical ensemble. The Hamiltonian is given by
eqs.(\ref{enerG})-(\ref{hamil}).

The classical partition function of the gas is then,
$$
Z(T,N,V)=\frac{1}{N!} \int\ldots\int \prod_{l=1}^{N}
\frac{{\rm d}^3 {\vec  p_{l}} \; {\rm d}^3 {\vec q_{l}}}{(2 \pi)^3} \;
 \; e^{- \frac{H}{T} } \; \;.
$$
\noindent It is convenient to introduce the dimensionless coordinates
${\vec r_{l}}=\frac{{\vec q_{l}}}{L}$.
The momenta integrals are computed straightforwardly. Hence, the
partition function factorizes as  the partition function of 
a perfect gas times the coordinate integral $Z_{coor}$.
 \begin{equation}\label{part}
Z=\frac{V^{N}}{N!} \left(\frac {m T}{2 \pi}\right)^{3 N/2}
 \; Z_{coor} \; ,
\end{equation}
where
\begin{equation}\label{coor}
Z_{coor}=\int\ldots\int \prod_{l=1}^{N}{\rm d}^3{\vec r_{l}} \; \;
e^{\eta  \, u_{P} +\frac{2 \pi}{3} \, \xi \, u_{N}} \; ,
\end{equation}
and
$$
u_{P} \equiv \frac {1}{N} \sum_{1 \leq i < j \leq N} 
\frac {1}{|{\vec r_{i}}-{\vec r_{j}}|} \quad , \quad
u_{N} \equiv \sum_{i=1}^{N}  r_{i}^2
 \; . 
$$
where $\eta$ and $\xi$ are defined by eq.(\ref{etaxi}).

The free energy  takes then the  form,
\begin{equation}\label{flib}
F = -T \log Z(T,N,V) = -N T \log\left[ \frac{ e V}{ N}
  \left(\frac{mT}{2\pi}\right)^{3/2}\right]  - T \; \log Z_{coor} \; , 
\end{equation}
We get for the  external pressure of the gas,
\begin{equation}\label{pres}
p = - \left(\frac{ \partial F}{\partial V}\right)_T = \frac{N T}{V} +
T \; \left(\frac{ \partial}{\partial V}\right)_T \log Z_{coor} \; .  
\end{equation}
Evaluating the volume derivative of $ Z_{coor} $ with the help of
eqs.(\ref{etaxi})  and (\ref{coor}) yields,
\begin{equation}\label{virial}
\frac{p V}{N T} = 1 +\frac{ <U_P> - 2 <U_{\Lambda}>} {3 N T}
\end{equation}
where
$$
{\cal U}_P = -G \sum_{i<j} \frac{m_i \; m_j}{|{\vec q}_i-{\vec q}_j|}
\quad \mbox{and} \quad
{\cal U}_{\Lambda} =- \frac{4\pi \, G \, \Lambda}{3} \,\sum_i m_i \;
q_i^2 \; . 
$$
Eq.(\ref{virial}) is the virial theorem for the gas of
self-gravitating particles in the presence of the cosmological constant.

\subsection{Mean field theory}

In the large $N$ limit the $3N$-uple integral in eq.(\ref{coor}) can
be approximated by a functional integral over the density (see
refs.\cite{gasn}). 
\begin {equation}
\label{intfonct}
Z_{coor}=\int D\rho(.)  \; {\rm d}b \; e^{-N \; {\cal S}[\rho(.)]}
\end{equation}
\noindent
with,
\begin {eqnarray}
\label{action}
{\cal S}[\rho(.)] &=& -\; \frac{\eta}{2}\int\frac{ {\rm d}^3{\bf x} \,
  {\rm d}^3{\bf y}} {|{\bf x}-{\bf y}|} \; \rho({\bf x})\; \rho({\bf y})
-\frac{2 \pi}{3} \; \xi \int{\rm d}^3{\bf x}  \; \rho({\bf x}) \; {\bf
x}^2 \nonumber\\ 
&&+ \int {\rm d}^3 {\bf x} \; \rho({\bf x}) \ln{\rho({\bf x} )}
+i \;  b \; [1-\int {\rm d}^3 {\bf x} \; \rho({\bf x})] \; \; ,
\end{eqnarray}
\noindent
and $b$ is the  Lagrange multiplier enforcing the normalization of the
density: 
\begin {equation} \label{normalisation}
\int {\rm d}^3 {\bf x}\;\rho({\bf x})=1   \; \; .
\end{equation}
The function $Z_{coor}$ thus becomes a functional integral over all
density configurations $\rho({\bf x})$.
Using eqs.(\ref{part}),(\ref{intfonct}) and (\ref{action}) the 
partition function of the system becomes
$$
Z=\int D\rho(.)  \; {\rm d}b \; e^{-\frac{1}{T} \, F[\rho(.)]} \; ,
$$
\noindent where 
\begin{equation} \label{freeen}
F[\rho(.)]=F_0+N \;T \; {\cal S}[\rho(.)]
\end{equation}
stands for the free energy of the gas with density $\rho$, while 
$$ 
F_0=-N T \ln{\left[\frac{e V}{N} \left(\frac{m T}{2 \pi}\right)^
{\frac{3}{2}}\right]}
$$ 
is the free energy of the perfect gas. One recognizes  the potential
energy (the gravitational energy due to 
the self-interaction of the particles and the gravitational energy due
to the interaction of the particles with the dark energy) in the
first line  of eq.(\ref {action}), while the second line contains the
entropy and the term fixing the total number of particles to be $N$.

\vspace{1 cm}

Since the free energy  becomes large in the thermodynamic limit
($N\gg1$), the functional integral  $Z_{coor}$ in eq.(\ref{intfonct}) is
dominated by the extrema of ${\cal S}[\rho(.)]$. Extremizing
this quantity with respect to the density $\rho({\bf x})$
yields the saddle point equation
\begin {equation} \label{sadeq}
\ln{\rho({\bf x})}=a+\eta \int
\frac{{\rm d}^3{\bf y}}{|{\bf y}-{\bf x}|} \; \rho({\bf y})
+\frac{2 \pi}{3} \; \xi \; {\bf x}^2 \; \; .
\end{equation}
\noindent  This equation defines the mean field approach. We set $ i b=
a + 1 $ and
\begin {equation} \label{phi} 
\rho({\bf x})=e^{\Phi({\bf x})} \; .
\end{equation}
\noindent 
We recognize in eq.(\ref{sadeq}) the gravitational potential
\begin{equation}   \label{potentiel}
V({\bf x})=-\frac{T}{m}[\Phi({\bf x})-a] \; .
\end{equation} 
\noindent
Inserting eq.(\ref{potentiel}) into eq.(\ref{phi}) yields 
the Boltzmann law for the density 
\begin {equation} \label{Boltzmann}
\rho({\bf x})=e^{a} \; e^{-\frac{m}{T}V({\bf x})}  \; ,
\end{equation}
\noindent 
containing the energy of a particle 
in the mean field gravitational potential $V$.
$e^{a}$  plays the role of a normalization constant.

Applying the Laplace operator to the saddle point equation (\ref{sadeq})
we find the differential equation
\begin {equation} \label{Poisson}
\nabla^2 \Phi({\bf x})+4 \pi \; \left( \; \eta \; e^{\Phi({\bf x})}
-\xi \right)=0
\;.
\end{equation}
Eq.(\ref{Poisson}) is {\bf identical} to the hydrostatic equilibrium
eq.(\ref{ecsd}). Therefore, hydrostatics and mean field are equivalent in the $
N \to \infty $ limit. 

It must be stressed that in the hydrostatic approach one has to
assume some local equation of state. Only a perfect gas equation
yields eq.(\ref{Poisson}). In the mean field approach one can compute the local
equation of state and {\bf demonstrate} that this is the one of a
perfect gas\cite{gasn,gas2,gas3}. 

\subsection{Spherically symmetric case}

For spherically symmetric configurations the mean field equation (\ref{ecsd})
becomes an ordinary non-linear differential equation. The
various thermodynamic quantities are expressed in terms of their solutions.

\begin{figure}[htbp]
\rotatebox{-90}{\epsfig{file=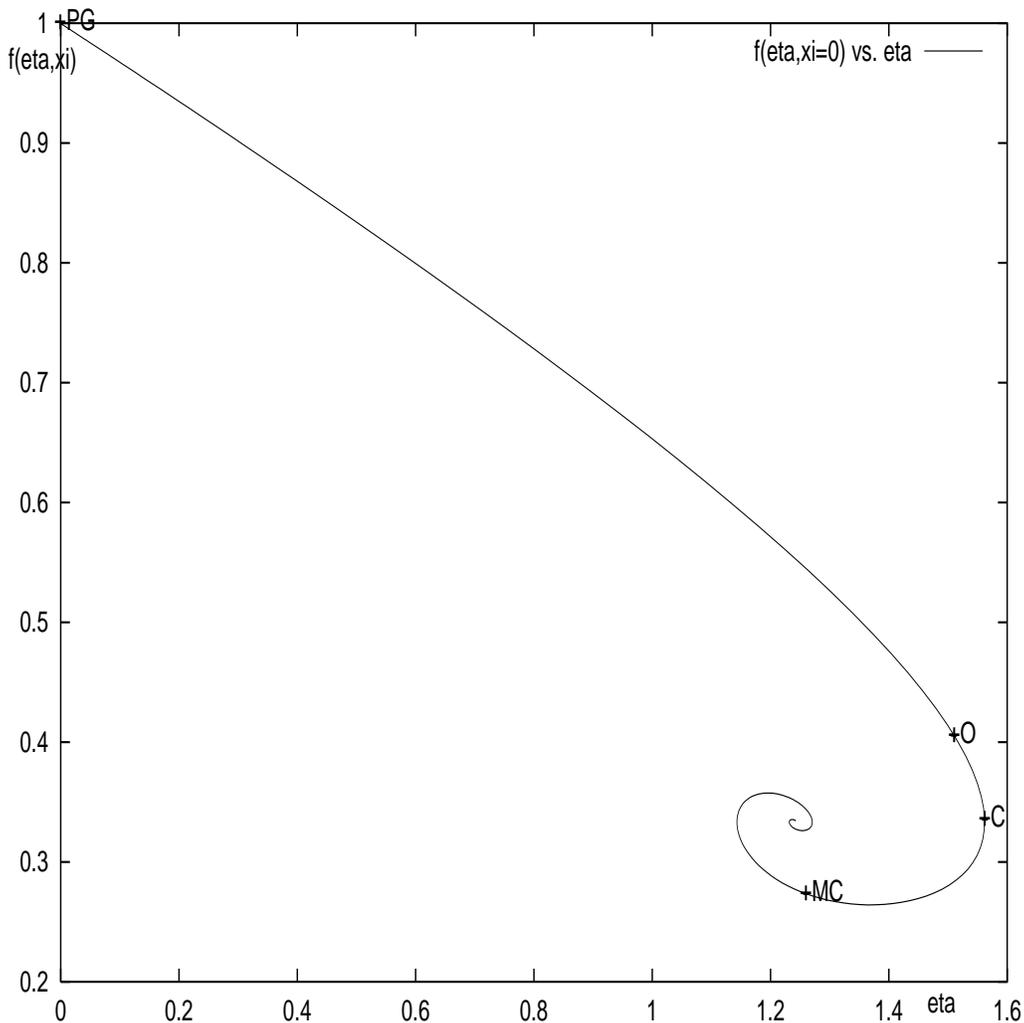,width=14cm,height=14cm}}
\caption{The pressure at the boundary of the self-gravitating gas without 
dark energy $ f(\eta,\xi=0) $ versus $\eta$.
PG is the perfect gas point, O is the point where the gas collapses
($ \eta^O=1.510\ldots $), C is the branch point where $ c_v $ diverges
($ \eta^C=1.561\ldots$), and MC is the collapse point in
the microcanonical ensemble ($ \eta^{MC}=1.259\ldots $). } 
\label{psansxi}
\end{figure}

\begin{figure}[htbp]
\rotatebox{-90}{\epsfig{file=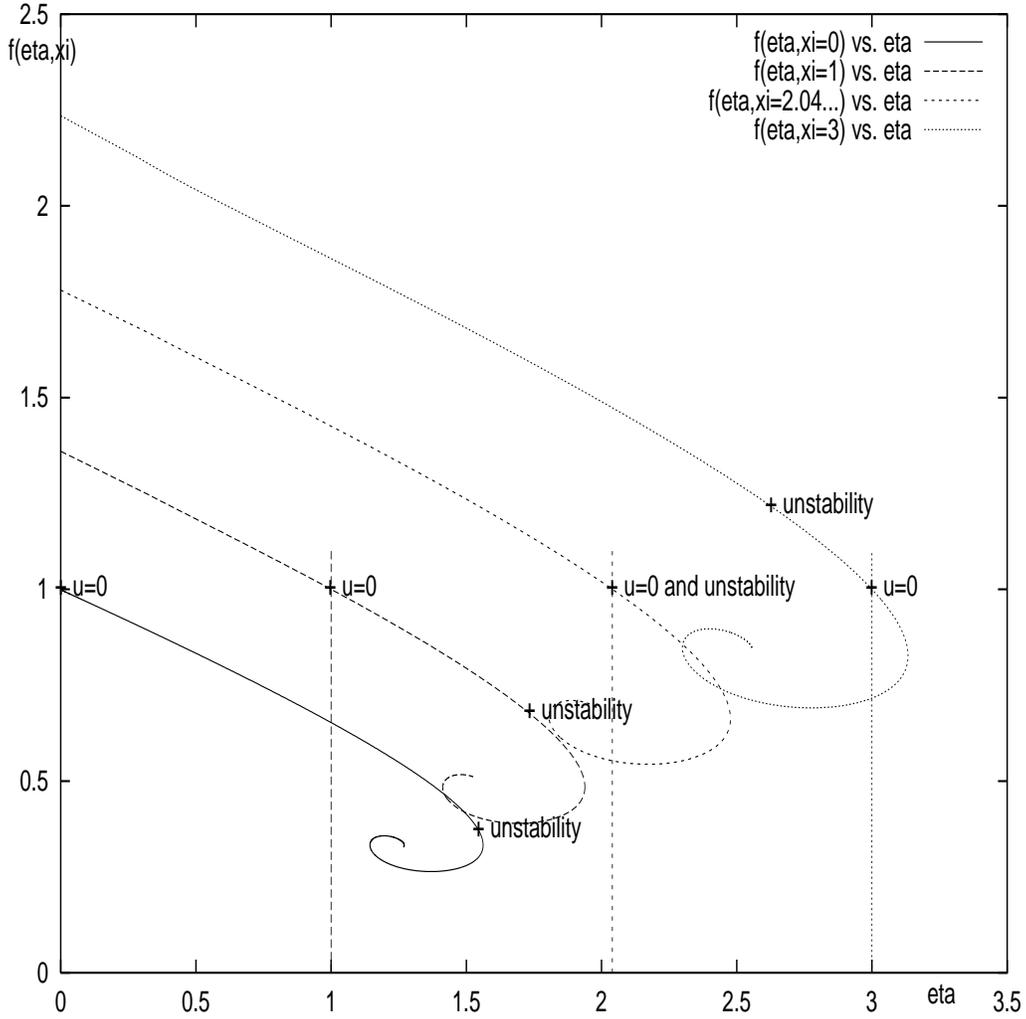,width=14cm,height=14cm}}
\caption{The density at the boundary  $f(\eta,\xi)$ versus $\eta$ for
$\xi=0,1,2.04\ldots,3$.
We insert on each graph the point of homogeneous density $u=0$ and the
point of unstability where the isothermal compressibility diverges.  
The gas is stable from $\eta=0$ till the point of unstability} 
\label{ps}
\end{figure}

\begin{figure}[htbp]
\rotatebox{-90}{\epsfig{file=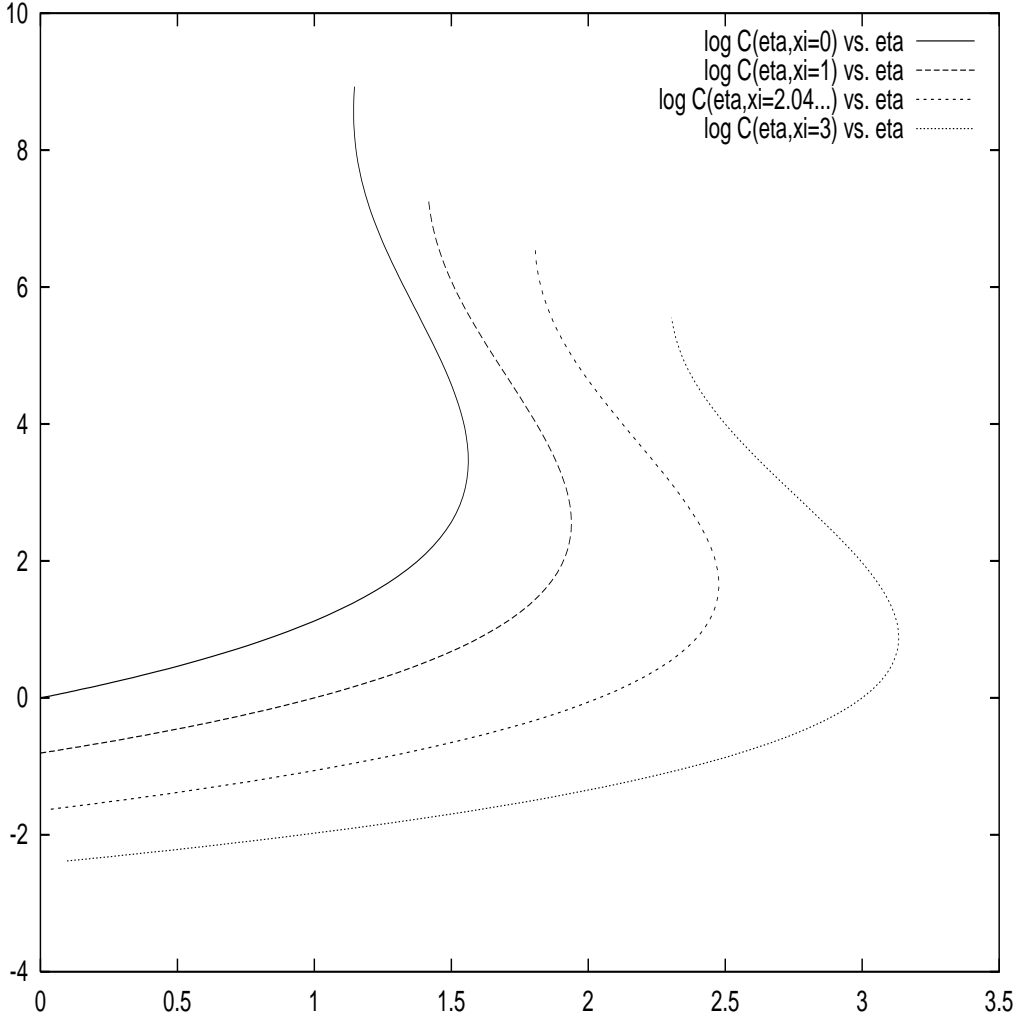,width=14cm,height=14cm}}
\caption{The logarithm of the contrast $ C(\eta,\xi) $ versus $\eta$ for
$\xi=0,1,2.04\ldots,3$. Notice that the contrast is unity for $\eta=\xi $.} 
\label{contrast}
\end{figure}

We consider here the spherically symmetric case where the mean field
equation (\ref{Poisson})  takes the form
\begin {equation} \label{Poissonrad}
\frac{{\rm d}^2 \Phi}{{\rm d}R^2}+\frac{2}{R} \frac {{\rm d} \Phi}{{\rm
d}R} +4 \pi \; \left(\eta \; e^{\Phi(R)}-\xi \right) = 0 \; .
\end{equation}
\noindent 
We work in a unit sphere, therefore the radial variable runs
in the interval $0 \leq R \leq R_{max} $,
$ R_{max} \equiv \left(\frac{3}{4 \pi}\right)^{\frac{1}{3}} $.
The density of particles  $\rho(R)$ has
to be normalized according to eq.(\ref {normalisation}).
Integrating eq.(\ref{Poissonrad}) from $R=0$ to $R=R_{max} $ yields,
\begin{equation} \label{emx}
\eta-\xi=-R_{max}^2 \; \Phi'(R_{max})
\end{equation}
\noindent 
Setting,
\begin{equation} \label{transf}
\Phi(R)=u(x)+\ln{\frac{\xi^{R}}{\eta^{R}}} \quad , \quad x=\sqrt
{3 \xi^{R}} \; \frac{R}{R_{max}} \; ,
\end{equation}
\noindent 
$\xi^{R}=\frac{\xi}{R_{max}}$ and $\eta^{R}=\frac{\eta}{R_{max}}$,
the saddle-point equation (\ref{Poissonrad}) simplifies as,
\begin{equation} \label{reducedeq}
\frac{{\rm d}^2 u}{{\rm d}x^2}+\frac{2}{x} \frac {{\rm d} u}{{\rm
d}x}+e^{u(x)}-1=0 \; .
\end{equation}
\noindent 
Let us find the boundary conditions for this  equation.
In order to have a regular solution at origin we have to impose
\begin{equation} \label{up0}
u'(0)=0 \; .
\end{equation}
\noindent
We find from eqs.(\ref{emx})-(\ref{transf}) for fixed values of $\eta$
and $\xi$,
\begin{equation} \label{emx2}
u'\left(\sqrt{3 \xi^{R}} \right)=-\frac{\eta^{R}-\xi^{R}}{\sqrt{3 \xi^{R}}}
 \; .
\end{equation}
Eqs.(\ref{up0}) and (\ref{emx2}) provide the boundary conditions for
the nonlinear ordinary differential equation (\ref{reducedeq}). In
particular, they impose the dependence of $u_0 \equiv u(0)$ on
$\eta^{R}$ and $\xi^{R}$. 

Using eqs.(\ref{phi}) and (\ref {transf}) we can express the
density of particles in terms of the solution of eq.(\ref {reducedeq})
\noindent
\begin{equation} \label{dens}
\rho(R)=\frac{\xi^{R}}{\eta^{R}} \; \exp[u(\sqrt
{3 \xi^{R}} \; \frac{R}{R_{max}})]\; .
\end{equation}
\noindent 
Recalling that the local pressure is $ P(R)=\frac{N \; T}{V} \;
\rho(R) $ and using eq.(\ref{dens}), the contrast $ C(\eta,\xi) $
between the pressure at the center and at the boundary can be  written as, 
$$
C(\eta,\xi)\equiv\frac{P(0)}{P(R_{max})}=e^{u_0-u(\sqrt
{3 \xi^{R}})} \; .  
$$
Using eqs.(\ref{phi}) and (\ref{transf}), the pressure at the boundary
is given by 
\begin{equation} \label{f}
P(R_{max})=\frac{N \; T}{V} \; f(\eta,\xi)=\frac{N \; T}{V} \;
\rho(R_{max})=\frac{N \; T}{V} \; \frac{\xi^{R}}{\eta^{R}} \; \exp[u(\sqrt {3
    \xi^{R}})]\; . 
\end{equation}
while the pressure at the origin takes the value
$$
P(0) = \frac{N \; T}{V} \; \frac{\xi^{R}}{\eta^{R}} \; e^{u_0} \; .
$$

\bigskip

We find from  eq.(\ref{reducedeq}) for small $x$ the 
expansion of  $u(x)$ in powers of $x$ with the result,
\begin{equation} \label{dl}
u(x)=u_0  + (1-e^{u_0}) \; \frac{x^2}{6}+e^{u_0} (e^{u_0}-1)   \;
\frac{x^4}{120} + {\cal O}(x^6)\; .
\end{equation}
\noindent where we imposed eq.(\ref{up0}). The value of $u_0$ follows
by imposing eq.(\ref{emx2}). This power expansion is well suited near
the center of the isothermal sphere [see sec IV.D]. 

Solving eq.(\ref{reducedeq}) with the boundary conditions $u(0)=u_0$,
$u'(0)=0$ and eq.(\ref{emx2})
yields the function $ u(R) $ for the whole range of values of $
\eta^R $ and $ \xi^R $ . We did that numerically using a fourth order
Runge-Kutta method. We plot in fig. \ref{psansxi} the density at the boundary
versus $\eta$ for $\xi=0$ and in fig. \ref{ps} the density at the boundary
versus $\eta$ for different values of $\xi$.
We plot in fig. \ref{contrast} the contrast $C(\eta,\xi)$ versus
$\eta$ for $\xi=0, \; 1, \; 2.04\ldots, \; 3$ .  As we can see from
fig. \ref{contrast}, the effect of the dark energy is
to {\bf decrease}  the value of the contrast in agreement with the repulsive
character of the dark energy. In particular, the contrast is unity for
$\eta=\xi $ . In this case the particle distribution is homogeneous
[see eq.(\ref{uig0}) and discussion there].

\bigskip

We  compute in what follows the thermodynamic quantities as functions of the
physical parameters $\eta$ and $\xi$.

\subsection{Free energy}

\begin{figure}[htbp]
\rotatebox{-90}{\epsfig{file=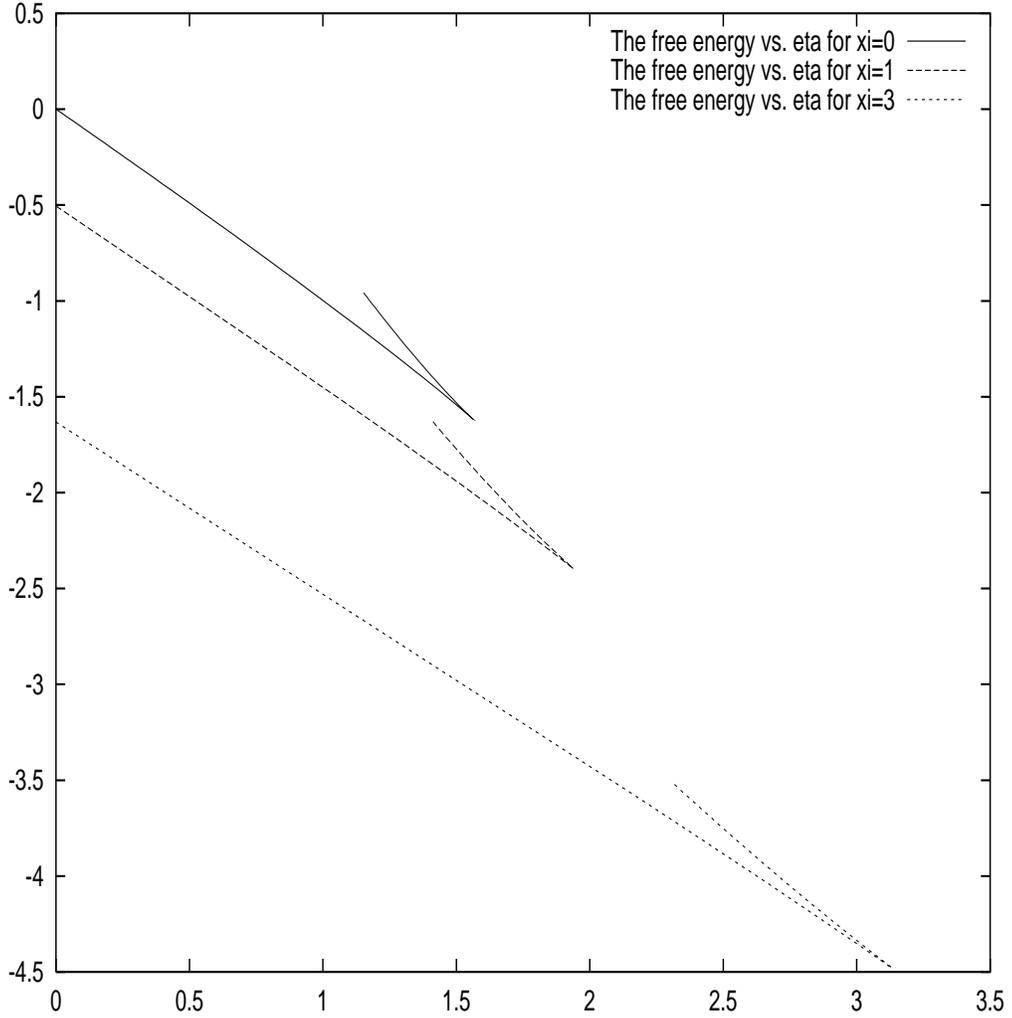,width=14cm,height=14cm}}
\caption{The free energy per particle divided by $T \; ,  \;
\frac{F(\eta,\xi)-F_0}{NT}$ versus $\eta$ for selected values of $\xi$.}
\label{freeener}
\end{figure}

\begin{figure}[htbp]
\rotatebox{-90}{\epsfig{file=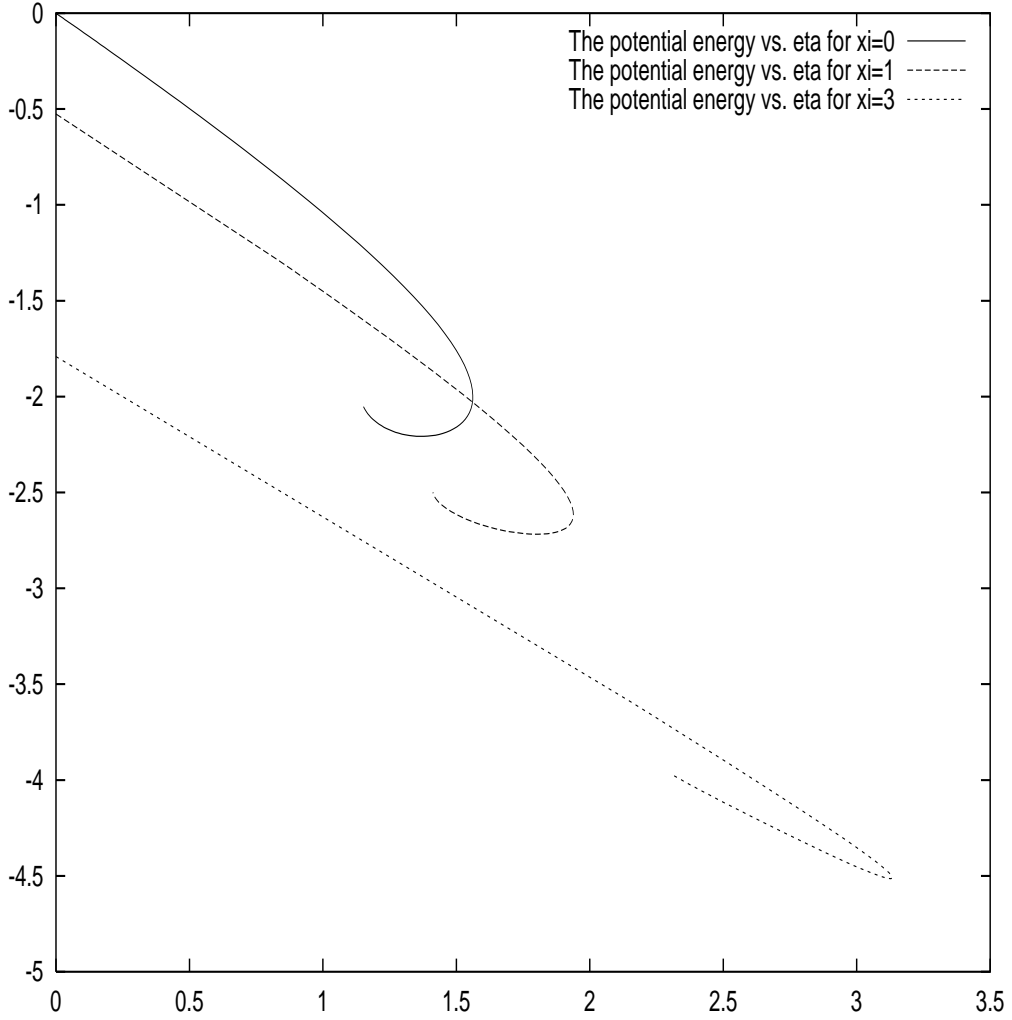,width=14cm,height=14cm}}
\caption{The potential energy per particle divided by $T \; ,  \;
  \frac{E_P(\eta,\xi)}{NT}$ versus $\eta$ 
  for  selected values of $\xi$.} 
\label{poten}
\end{figure}

\begin{figure}[htbp]
\rotatebox{-90}{\epsfig{file=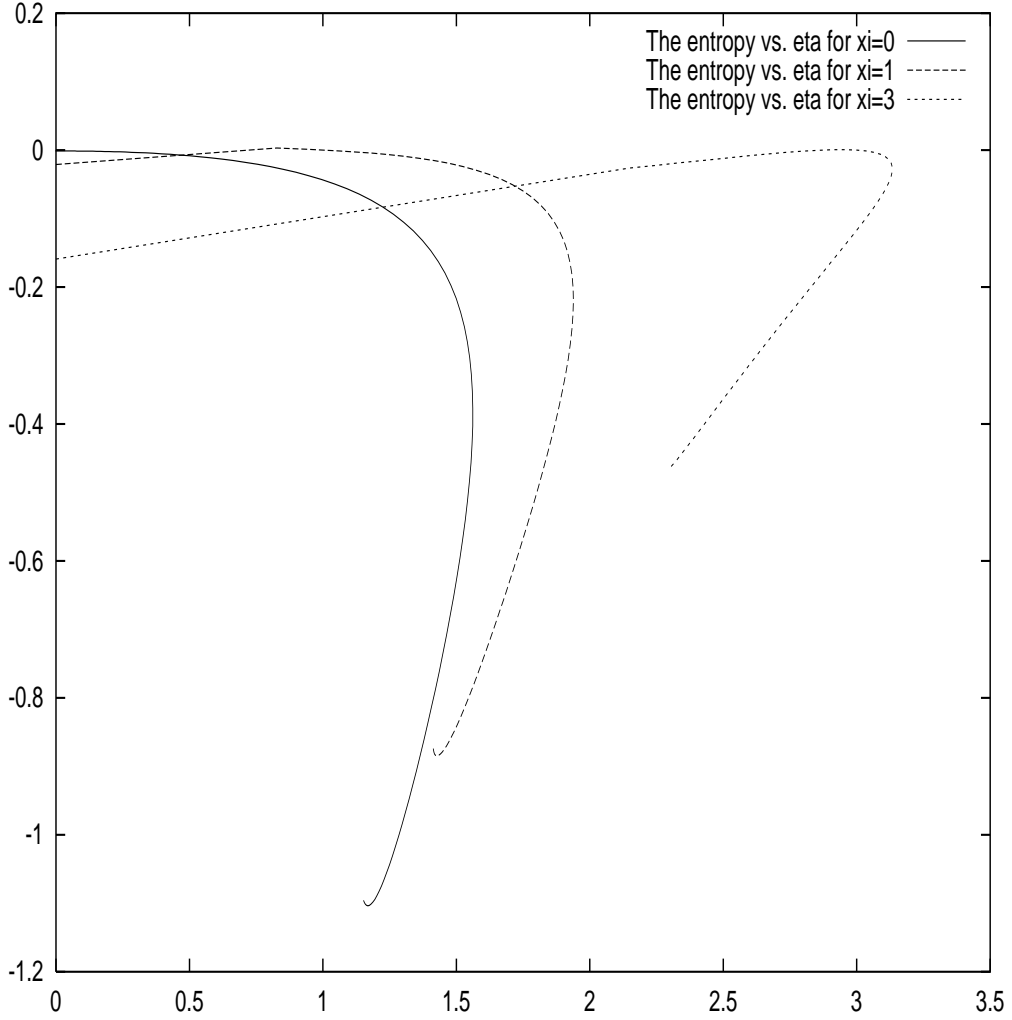,width=14cm,height=14cm}}
\caption{The entropy per particle $\frac{S(\eta,\xi)-S_0}{N}$ vs. $\eta$ for
  different  fixed values of $\xi$.} 
\label{entropy}
\end{figure}

We start by computing the free energy.
Using eqs.(\ref {action})-(\ref{freeen}) we find 
\begin {equation} \label{free1}
\frac{F-F_0}{N T}=\frac{1}{2}\left[a+\int {\rm d}^3{\vec x} \;
\Phi({\vec x}) \; e^{\Phi({\vec x})}-\frac{2 \pi}{3} \; \xi \; \int
{\rm d}^3{\vec x} \; x^2 \; e^{\Phi({\vec x})}\right] \;.
\end{equation}
\noindent
In order to express the Lagrange multiplier $a$ as functions of the 
physical parameters $\eta$ and $\xi$ we use eq.(\ref {sadeq}) where we
can now integrate over the angles with the result,
\begin {equation} \label{intangle}
\Phi(R)=a+4 \pi\eta\left[\frac{1}{R}\int_{0}^{R} {\rm d}R'
\; R^{'2} \; e^{\Phi(R')} +\int_{R}^{R_{max}}{\rm d}R' \;
R' \; e^{\Phi(R')} \right] + \frac{2 \pi}{3} \; \xi \;R^2
\end{equation}
\noindent
We call $ f(\eta^{R},\xi^{R}) $ the density of particles at the
boundary $R=R_{max}$ [see eq.(\ref{f})],   
\begin {equation} \label{densper}
f(\eta^{R},\xi^{R}) \equiv e^{\Phi(R_{max})}=\frac{\xi^{R}}{\eta^{R}} \;
e^{u(\sqrt{3 \xi^{R}})} \; . 
\end{equation}
\noindent Setting $R=R_{max}$ in eq.(\ref{intangle}) yields for the
Lagrange multiplier 
\begin{equation} \label{Lagrange}
a=\ln f(\eta^{R},\xi^{R}) -\eta^R-\frac{\xi^{R}}{2} \; .
\end{equation}
where we used the normalization of the densities  eq.(\ref
{normalisation}) 

\noindent
Inserting eq.(\ref {Lagrange}) into the free energy (\ref {free1}), we find 
\begin{equation} \label{free2}
\frac{F-F_0}{N T}=\frac{1}{2} \; \left[\ln
f(\eta,\xi)-\eta^R-\frac{\xi^R}{2}\right]+ 2 \, \pi \int_0^{R_{max}} R^2 \; dR 
\; \left[ \Phi(R)-\frac{2 \pi}{3}\; \xi \; R^2 \right] \; e^{\Phi(R)}
\; .
\end{equation}
\noindent
These integrals  are computed in appendix A with the result
\begin {eqnarray} \label{free}
&&\frac{F-F_0}{N T}=3[1-f(\eta^{R},\xi^{R})]-\eta^R+ \left( \frac{2 \,
\xi^R}{\eta^R} + 1 \right) \ln{f(\eta^R,\xi^R)}  \nonumber\\
&&+\frac{\xi^R}{2}\left(1-\frac{4}{5}\;\frac{\xi^R}{\eta^R}\right)
-8 \, \pi\;\frac{\xi^R}{\eta^R} \; \int_0^{R_{max}}{\rm d}R \;
R^2 \;\Phi(R) \; .
\end{eqnarray}
We plot in fig. \ref{freeener} the free energy as a function of
$\eta$ for fixed $\xi$. It  
exhibits two Riemann sheets. We call  $ \eta_C(\xi)$ the position of the
 branch point. The curves have the same qualitative behaviour as for $\xi=0$
except that the branch point  $\eta_C(\xi)$ increases with $\xi$.
 The dark energy moves the plot of the free energy to lower
values. It is a consequence  of the negative contribution of  the dark
energy to the energy [eq.(\ref{enerG})] and the free energy
[eq.(\ref{free1})].   

\bigskip

When matter dominates over dark energy we have $ N \; m \gg  V \;
\Lambda$ and therefore according to  eq.(\ref {etaxi}) $\eta \gg
\xi$.  In such limit the free energy becomes,
$$ 
\frac{F-F_0}{N T}\buildrel{\eta \gg \xi}\over=\ln f(\eta^R,\xi^R)
-\eta^R+3\left[1-f(\eta^R,\xi^R)\right] \; . 
$$ 
\noindent
We recognize here the free energy of a self-gravitating gas with 
$N$ particles of mass $m$ as it must be (see ref.\cite{gasn}).

\subsection{Energy and entropy}

We compute here the gravitational energy of the particles.
The gravitational potential is the sum of the self-gravity of the
particles plus their interaction with the dark energy [see
eq.(\ref{enerG})].  Equivalently, we find for  potential energy from
eqs.(\ref{potentieltot}), (\ref{etaxi}), (\ref{vincu2}) and (\ref{potentiel}),
$$
E_P=\frac{1}{2}\;\int \rho({\vec q}) \; V_{self}(\vec{q}) \;d^3{\vec q}
+ \int \rho({\vec q}) \; V_{dark}(\vec{q}) \;d^3{\vec q}=
NT \left[\frac{a}{2}-\frac{1}{2} \int {\rm d}^3{\vec x} \;
\Phi({\vec x}) \; e^{\Phi({\vec x})}-\frac{2 \pi}{3} \; \xi \; \int
{\rm d}^3{\vec x} \; x^2 \; e^{\Phi({\vec x})}\; \right] \; ,
$$
where we use the dimensionless coordinates $\vec{x}=\frac{\vec{q}}{L}$
and $a$ is given by eq.(\ref{Lagrange}).

We compute these integrals for the spherically symmetric case in
appendix A with the result,
\begin {eqnarray} \label{energy}
\frac{E_P}{N T}&=&3[f(\eta^{R},\xi^{R})-1] -\frac{11}{8} \;\xi^R\left( 1 -
\frac{2}{5} \;\frac{\xi^R}{\eta^R}\right) -\frac{11}{4}\;\frac{\xi^R}{\eta^R} 
\; \ln{f(\eta^R,\xi^R)}+ \nonumber\\
&&+ 11 \; \pi \; \frac{\xi^R}{\eta^R} \;
\int_0^{R_{max}}{\rm d}R \; R^2 \;\Phi(R) \; .
\end{eqnarray}
We plot the gravitational energy as a function of $ \eta $ for fixed
values of $ \xi $ in fig. \ref{poten}. We see a two-sheeted structure
with the branch point at $\eta_C(\xi)$.
The dark energy pushes the potential energy to lower
values. As discussed before for the free energy, it is a consequence
of the repulsive character of the dark energy. 

\bigskip

We now compute the entropy as $ S = [E - F]/T $ using eqs.(\ref{free}) and
(\ref{energy}) for the free energy and the energy, respectively,
\begin{eqnarray} \label{entr}
 \frac{S-S_0}{N}&=&6\left[f(\eta^R,\xi^R)-1\right]
+\eta^R  -\frac{15}{8}\xi^R+\frac{19}{20}\frac{\xi^R}{\eta^R}\xi^R
-\left(1 + \frac{19}{4}
\;\frac{\xi^R}{\eta^R} \right)\ln{f(\eta^R,\xi^R)}
+ \nonumber\\
&&+19 \, \pi \;\frac{\xi^R}{\eta^R} \; \int_0^{R_{max}}{\rm d}R \; R^2 \;
\Phi(R) \; . 
\end{eqnarray}
We plot the entropy as a function of $ \eta $ for fixed values of $
\xi $ in fig. \ref{entropy}. We see again a two-sheeted structure
with the branch point at $\eta_C(\xi)$.

\subsection{Stability of the gaseous phase}

\begin{figure}[htbp]
\rotatebox{-90}{\epsfig{file=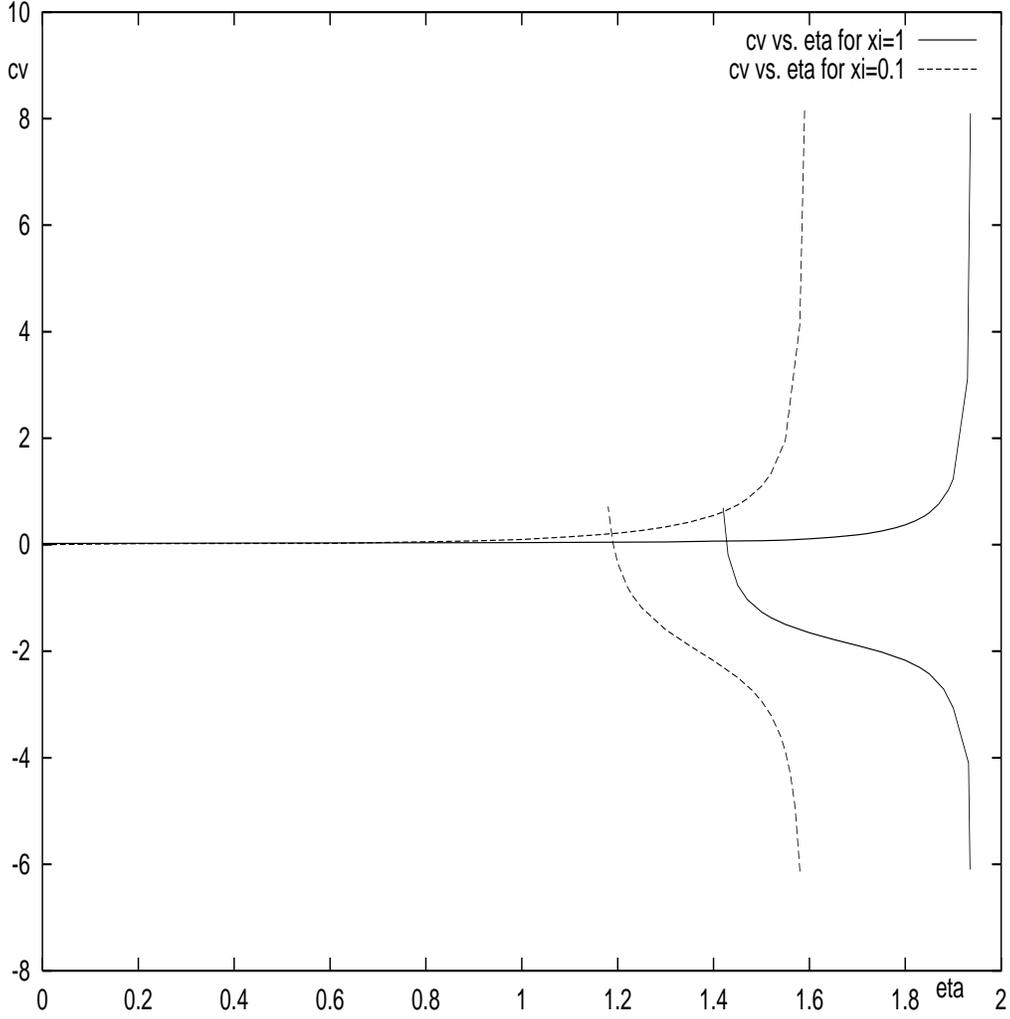,width=14cm,height=14cm}}
\caption{The specific heat per particle at constant volume, $c_v$
  vs. $\eta$ for  fixed $\xi=0.1,1$.} 
\label{cv}
\end{figure}

\begin{figure}[htbp]
\rotatebox{-90}{\epsfig{file=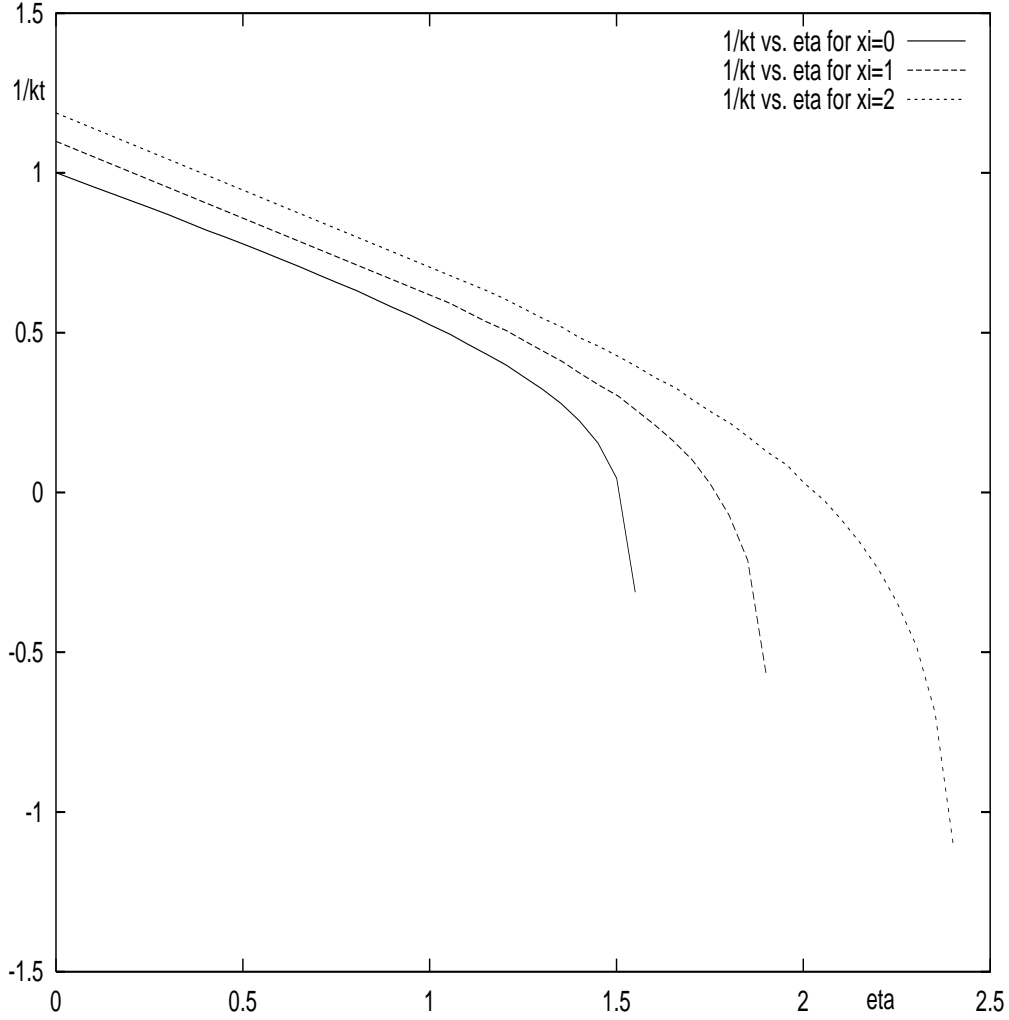,width=14cm,height=14cm}}
\caption{The inverse of the isothermal compressibility vs. $\eta$ for 
fixed $\xi=0,1$ and $2$.} 
\label{kt}
\end{figure}

In the precedent paragraphs we present the mean field theory of the
selfgravitating gas in the presence of dark energy. 
Mean field theory gives exactly the physical quantities of the gaseous
phase in the thermodynamic limit 
$$
N \to \infty \quad , \quad L \to \infty \quad , \quad \Lambda \to 0 \;
, 
$$
with 
$$
\frac{N}{L} = \mbox{fixed} \quad , \quad \Lambda \, L^2 = \mbox{fixed}
\; .
$$
\noindent But it does not tell anything directly about the stability of the
gaseous phase. 

Necessary conditions for stability in the canonical ensemble are
positive specific heat $c_v$ and compressibility
$\kappa_T$\cite{hidro,gasn,gas2}. We compute below  $c_v$  per particle and
$K_T$ starting from their definitions, 
\be\label{defcv}
c_v=\frac{T}{N} \; \left( \frac{\partial S}{\partial T} \right)_{V}
\quad ,  \quad 
K_T=-\frac{1}{V}  \; \left( \frac{\partial V}{\partial P}
\right)_{T} \; .
\ee
The point of instability of the self-gravitating gas in absence of
dark energy is $\eta^0=1.510\ldots$ (see fig. \ref{psansxi}) where the
compressibility diverges\cite{gasn}. At this point 
the isothermal compressibilities has a simple pole 
and changes sign. It is also the point where the speed
of sound at the center of the sphere becomes imaginary. Small
density fluctuations grow exponentially in time instead of exhibiting
oscillatory propagation. Such a behavior clearly leads to instability
and collapse.

\bigskip

In order to compute the specific heat for nonzero $\xi$ 
we introduce the function
\begin{equation} \label{g}
g(x,u_{0})=\left(\frac{\partial u}{\partial u_{0}}\right)_{x}(x,u_{0}) \; .
\end{equation}
\noindent
Using eqs.(\ref{reducedeq}) and (\ref{up0}) we find that this function
$g$ obeys to the second order differential equation 
$$
\left(\frac{\partial^2 \; g}{\partial x^2}\right)_{u_{0}}+\frac{2}{x}
\; \left(\frac{\partial g}{\partial x} \right)_{u_{0}}+
e^{u(x,u_{0})}\; g(x,u_{0})=0 \; ,
$$
\noindent
with the boundary conditions
$$
g(0,u_{0})=1 \;\;\;\;\;\;\;\;   \left(\frac{\partial g}{\partial
  x}\right)_{u_{0}}(0,u_{0})=0 
$$ 
\noindent
$c_{v}$ is computed in the Appendix B with the result,
\begin{eqnarray} \label{cv2}
&& c_{v}=3 \; (\eta^{R}-\xi^{R}) \; f -\frac{57}{8} \; \frac{\xi^{R}}{\eta^{R}}
\; \ln{f} -\frac{3}{2} \; \eta^{R}+\frac{57}{40}\; \frac{\xi^{R}}{\eta^{R}} \;
\xi^{R}+ \frac{57}{2}\pi \;\frac{\xi^{R}}{\eta^{R}}\; 
\int_{0}^{R_{max}} 
{\rm d}R \; R^2 \;\Phi(R,u_{0})  \nonumber\\
&& -\left[\eta^{R} \; \left( \frac{\partial u_{0}}{\partial \eta^{R}} \right)+
\xi^{R} \; \left( \frac{\partial u_{0}}{\partial \xi^{R}} \right) \right] \;
\left[ \left( 6 \; f-1- \frac{19}{4}\frac{\xi^{R}}{\eta^{R}} \right)  \;
  g\left(\sqrt{3 \; \xi^{R}}, 
u_{0}\right)   \right. \nonumber\\
&& \left. +\frac{57}{4} \; \frac{\xi^{R}}{\eta^{R}} \;
\frac{1}{(3 \xi^{R})^{\frac{3}{2}}} 
\int_{0}^{\sqrt{3 \; \xi^{R}}} {\rm d}x \; x^2 \; g(x,u_{0}) \right] 
\end{eqnarray}
We see in fig.\ref{cv} that the specific heat at constant volume
versus $\eta$ for a fixed $\xi$ is positive in the first branch till
the branch point 
($\eta=\eta^C(\xi)$), where  the specific heat tends to infinite and
changes its sign. $c_v$ is negative in the second branch, therefore the
second branch of the gaseous phase can not be stable in the canonical ensemble.

\bigskip

Using eqs.(\ref{f}) the dimensionless isothermal compressibility takes
the form,
\begin{equation} \label{kappat}
\kappa_T\equiv\frac{NT}{V} \; 
K_T = \frac{1}{f(\eta^R,\xi^R)+\frac{\eta^R}{3}
  \frac{\partial f}{\partial 
\eta^R}-\frac{2 \xi^R}{3} \frac{\partial f}{\partial\xi^R}}
\end{equation}
\noindent We plot in fig.\ref{kt} $ \frac{1}{\kappa_T} $ versus $ \eta
$ for  fixed $ \xi $. The isothermal compressibility tends to infinite and
becomes negative at the point $\eta=\eta^0$ indicating the collapse of
the gas. This unstability happens for values of $\eta^0(\xi)$ {\bf
  smaller} than $\eta^C(\xi) $ (see Table 1). 
Thus the positivity of the isothermal compressibility is a stronger
condition of stability than the positivity of the specific heat at
constant volume. For $\xi=0$ Monte Carlo simulations shows that the
gas collapse in a very dense phase at the point of divergence of the
isothermal compressibility \cite{gas2}.  
Mean field indicates that the gas in the presence of dark energy is
stable from $\eta=0$ till the point  $\eta=\eta^0(\xi)$. 

\bigskip

\begin{tabular}{|l|l|l|}\hline
$ \xi $ & $\eta^{0}(\xi)$ & $\hspace{0.5cm} \eta^C(\xi) $
\\ \hline $ 0 $ & $ 1.510... $ & \hspace{0.3cm}
$ 1.561... $  \hspace{0.3cm} \\
\hline $ 1 $ & $ 1.76... $ & \hspace{0.3cm}
$ 1.938... $  \hspace{0.3cm} \\
\hline $ 1.7 $ & $ 1.88... $ & \hspace{0.3cm}
$ 2.282... $  \hspace{0.3cm} \\
\hline $ 2.04... $ & $ 2.04... $ & \hspace{0.3cm}
$ 2.476... $  \hspace{0.3cm} \\
\hline $ 3 $ & $ 2.58... $ & \hspace{0.3cm}
$ 3.133... $  \hspace{0.3cm} \\
\hline $ 4.171... $ & $ 3.46... $ & \hspace{0.3cm}
$ 4.171... $  \hspace{0.3cm} \\
\hline $ 6 $ & $ 4.92... $ & \hspace{0.3cm}
$ 6.307... $  \hspace{0.3cm} \\
\hline
\end{tabular}

\bigskip

{TABLE 1. $\eta^0(\xi)$ and $\eta^C(\xi)$ for different values of $\xi$}

\section{Physical Picture}

\begin{figure}[htbp]
\rotatebox{-90}{\epsfig{file=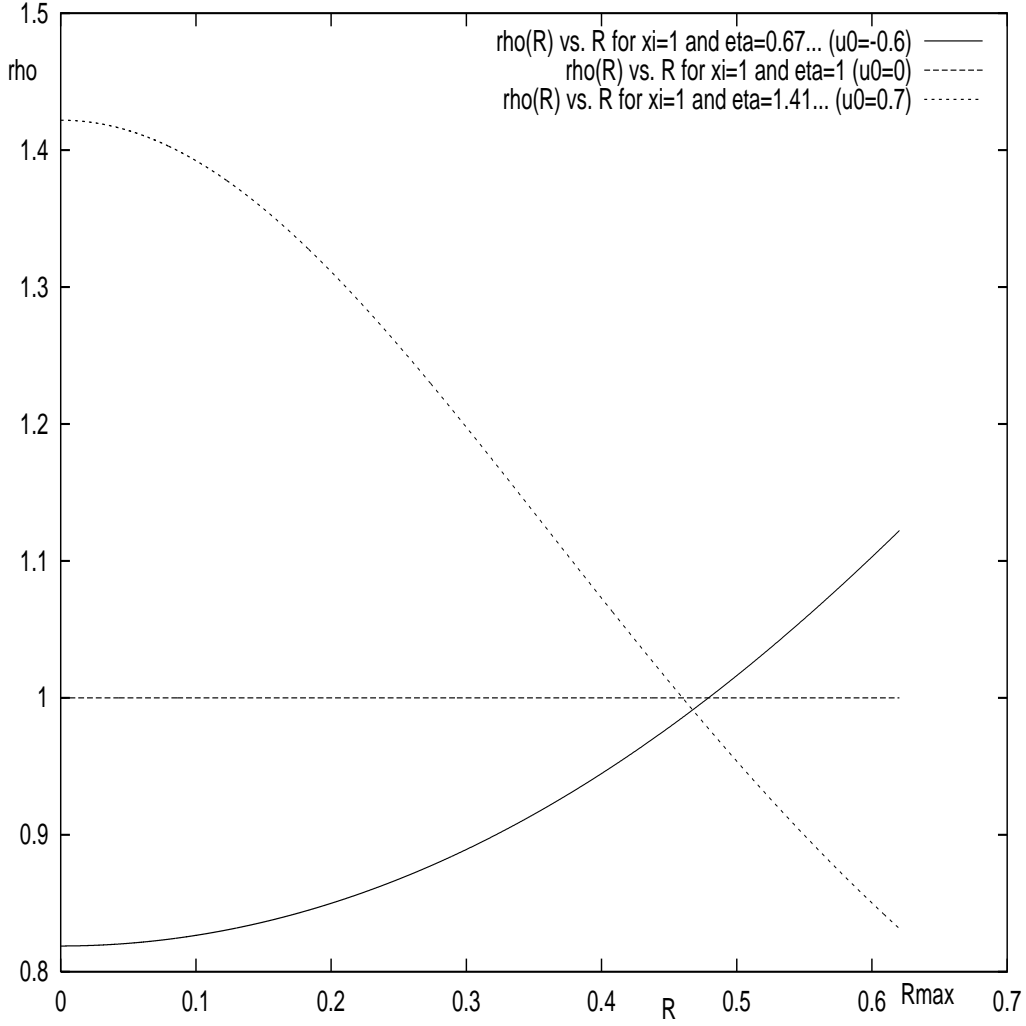,width=14cm,height=14cm}}
\caption{The densities $\rho(R)$ vs. the radial coordinate $R$ 
for $\xi=1$ and $\eta = 0.67\ldots, \; 1 $ and $ 1.41\ldots$.}
\label{rho1}
\end{figure}

\begin{figure}[htbp]
\rotatebox{-90}{\epsfig{file=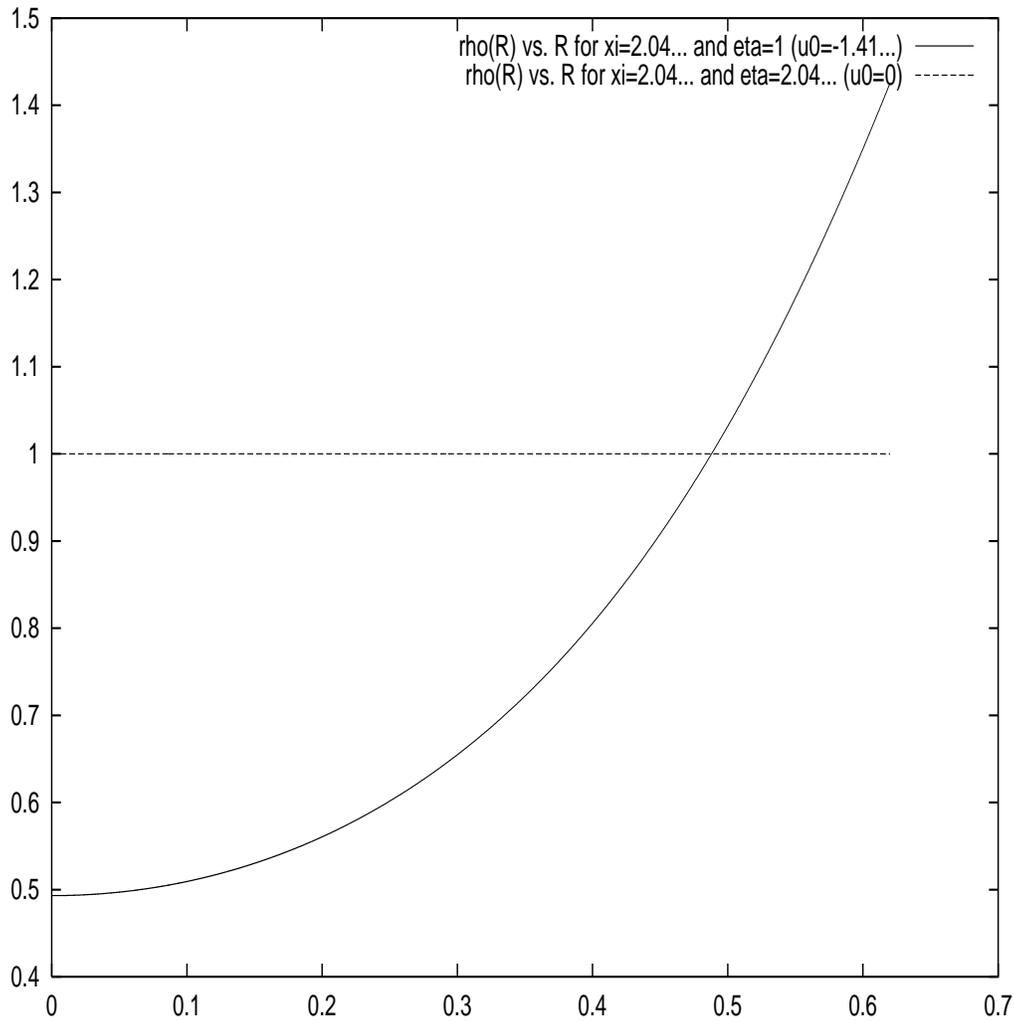,width=14cm,height=14cm}}
\caption{The densities $\rho(R)$ vs. the radial coordinate $R$ 
for $\xi=2.04\ldots$. The density increases with $R$ in the
first plot. The density is  uniform in the second plot and
coincides with the point of instability $\eta=\eta^0(\xi)$ .} 
\label{rho2p04}
\end{figure}

\begin{figure}[htbp]
\rotatebox{-90}{\epsfig{file=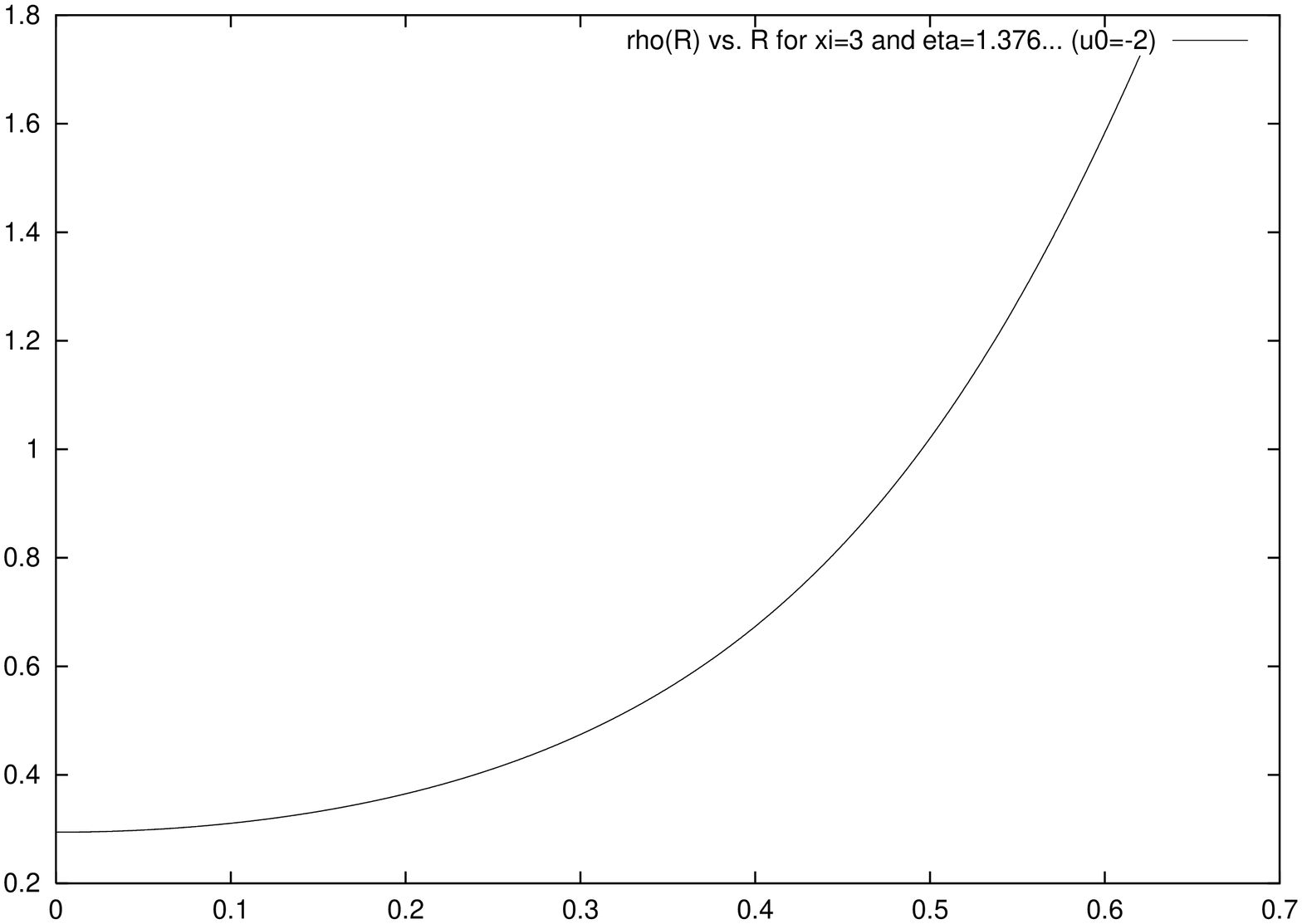,width=14cm,height=14cm}}
\caption{The density $\rho(R)$ vs. the radial coordinate $R$ is an
 increasing function  for $\xi=3$.}
\label{rho3}
\end{figure}

Let us first recall the properties of the self-gravitating gas in
absence of dark energy ($\xi=0$) \cite{gasn}. The self-gravitating
system  can be in one of 
two phases: gaseous or highly condensed. The pressure at the
boundary $ f(\eta,\xi=0) $ as a function
of $\eta$ is given in fig.\ref{psansxi} in units of $ \frac{N \; T}{V}
$. The function $ f(\eta,\xi=0) $ is calculated from mean field theory
and describes the gaseous phase. $ f(\eta,\xi=0) $  has two
Riemann sheets in the $\eta$-plane. The mean field theory gives exactly the 
physical quantities in the thermodynamic limit. The stability of the
gas phase follows by studying the speed of sound inside the sphere and
was moreover investigated by Monte Carlo simulations \cite{gasn,gas2}.
The gaseous phase is stable only in one piece of the curve $
f(\eta,\xi=0) $. This zone of stability depends of the statistical ensemble. 
In the first branch (PG-C) $ f(\eta,0) $ monotonically decreases for
increasing $\eta$ as shown in fig. \ref{psansxi}. That is, for decreasing
temperature or increasing 
density $N/L$. The specific heat at constant volume is positive and the
gaseous phase can be stable in the microcanonical and in the canonical
ensemble. In fact, the gaseous phase is stable in the canonical
ensemble from the point PG ($\eta=0$, perfect gas) till the point
O ($\eta^O=1.510...$) where the speed of sound at the center becomes
imaginary and the isothermal compressibility tends to infinite and changes 
its sign \cite{gasn,gas2}. Monte-Carlo simulations show that the gas
collapses at this point to a extremely condensed phase\cite{gasn}. This
is before the point C ($\eta^{C}=1.561\ldots$) where $
f(\eta,0) $  exhibits a vertical 
slope. C is the branch point of the function $ f(\eta,0) $.
The determinant of small fluctuations vanishes and the specific
heat at constant volume becomes infinite at C\cite{gasn,gas2}. In the
second branch (C-MC) the specific heat at constant volume is
negative. Then, the gaseous phase cannot be stable in the canonical
ensemble in this second branch  but it can be stable in the microcanonical
ensemble in (C-MC). In the microcanonical ensemble the gaseous phase is stable
from PG to MC and collapses in a condensed phase at the point
MC ($\eta^{MC}=1.259\ldots$)\cite{gasn,gas2}. 
Beyond the point MC $f(\eta,0)$ develops a nice spiral form, but 
the gas is unstable in all three statistical ensembles. 

That is, solving just the mean field equation does not tell us whether
the gas is stable. One has to compute in addition the isothermal
compressibility (or the local speed of sound) and the specific heat at
constant volume.  

In the gaseous phase the free energy has a minimum for the density
$\rho({\vec x})$, solution of the saddle point equation (\ref
{sadeq}). This density is the most probable distribution
which becomes absolutely certain in the infinite $N$  limit. All
thermodynamic quantities follow from this density $\rho({\vec x})$.  
In the condensed phase, the system becomes extremely dense with all
particles on the top of each other. The solution of the saddle point equation
 ceases to be a minimum of the free energy and the mean field approach
 then fails to describe the condensed phase. The condensed phase has
 been found by Monte-Carlo methods in ref.\cite{gasn}.

\bigskip

Let us discuss the properties of the self-gravitating gas in the
presence of dark energy. 

\subsection{The Triple Point}

\noindent The effects of self-gravitation and dark
energy are  opposed. Self-gravitating forces are attractive while 
dark energy produces repulsion.  If the parameters $\eta$ and $\xi$,
which determine the strength of the gravitation and 
dark energy, respectively, are equal, a exactly homogeneous sphere
$\rho(R)=1$ is a  solution (see below). In such special case the
self-gravitating gas behaves as a perfect gas (with $PV=NT$
everywhere).  

Such homogeneous solution is present for all $ \eta = \xi \geq
0$. There is a peculiar point where the homogeneous solution coincides with 
the point of unstability $\eta=\eta^0(\xi)$.
The point $\eta=\eta^0(\xi)$ is defined as the point where the isothermal compressibility
diverges and changes its sign.
That is, the two conditions
 $$
f(\eta=\xi,\xi)=1 \quad  \mbox{and} \quad \eta^0(\xi)=\xi
$$ 
\noindent define an unique `triple point' $\eta=\xi=\xi^0=2.04\ldots $.

\subsection{The critical point as a function of $\xi$}

\noindent In fig. \ref{ps} we
plot the pressure at the boundary versus $\eta$ for different  values of $\xi$.
We see that the plots for non zero $\xi$ exhibit a form analogous to
the plot for $\xi=0$. The effect of the dark energy is to conformally translate
the curve to higher pressure and higher $\eta$. Notice that the value
of the point of unstability $\eta^0(\xi)$ increases for increasing $\xi$.
For $ \xi > 0 $ the gas collapses at a higher density
$\frac{N}{L}$ for a given temperature and for a lower temperature for
a given density $\frac{N}{L}$ than for $\xi=0$. That is, the domain of
stability of the gas {\bf increases} for increasing $\xi$. The dark
energy has an antigravity effect that disfavours the collapse pushing
the particles towards the boundary of the sphere.  

\subsection{The particle  density $\rho(R)$}

We study the density $\rho(R)$  as a function of the radial coordinate $R$. The
density $\rho(R)$ of the self-gravitating gas in absence of dark
energy ($\xi=0$) is always a decreasing function of  $R$. This follows from the
attractive character of gravitation. At the point PG($\eta=0$)  [see
fig. \ref{psansxi}] the gas is ideal and
homogeneous, and $\rho(R)=1$. The farther we move from the point
PG($\eta=0$) in the $\xi=0$ case, the more inhomogeneous and faster
decreasing is the density $\rho(R)$. 

The $R$-dependence of the density
$\rho(R)$ of a self-gravitating gas in presence of dark energy is more
involved because the dark energy opposes to the attraction by gravity. 
The functions $\rho(R)$ and $u(x)$ have the same qualitative dependence in
$R$ since they are related by exponentiation [see eq. (\ref{dens})].

The behaviour  at the center of the sphere ($R=0$) is
governed by the sign of $u_0$ since $u'(0)=0$ [eq. (\ref{up0})] and we find
from eq.(\ref{dl}) that $ u''(0)= \frac{1}{3} \! 
\left(1-e^{u_0} \right)$. Hence,
$$
\mbox{sign}[ u''(0) ] = -\mbox{sign}[ u_0 ] \; .
$$
The behaviour at the boundary of the sphere
($R=R_{max}$) is governed by the sign of $\eta-\xi$ since according to
eq.(\ref{emx2})
$$
\mbox{sign}[ u'(R_{max}) ] = -\mbox{sign}[ \eta^{R}-\xi^{R} ] \; .
$$

\medskip

The particle density exhibits for the stable solutions (namely, for
positive $\kappa_T$) one of the following behaviours:

\begin{itemize}

\item{\bf decreasing}: $u_0>0$ and $\eta>\xi$. The density $\rho(R)$ decreases
 from the center of the sphere till the boundary. This is the only
 behaviour present when $ \xi = 0 $.

\item{\bf increasing}: $u_0<0$ and $\eta<\xi$. The density $\rho(R)$
  increases from the center of the sphere till the boundary. 
 \end{itemize}

More precisely, we find for the stable solutions the following
properties in the interval $ 0 \leq x \leq \sqrt{3 \xi^{R}} $:

\begin{itemize}

\item when $\eta>\xi, \; u(x)$ is positive and $ u'(x) < 0$.

\item when $\eta<\xi , \; u(x) $ is negative  and $ u'(x) > 0$.

\end{itemize}

\bigskip

\noindent The case {\bf $\eta=\xi$} is particularly interesting,
because we have $u'(0)=0$ and $u'\left(\sqrt{3 \xi^{R}} \right)=0$. 
This  implies  that 
\begin{equation} \label{uig0}
u(x)=0 \; , 
\end{equation} 
\noindent
is a solution of eq.(\ref{reducedeq}). However, there exist another
solution $u \neq 0$ for $\eta=\xi\geq 1.638\ldots$, but this solution
is never stable since $ \kappa_T < 0 $ for it. For the $u=0$ solution,
$\rho(R)=1$ and the gas is homogeneous. The self-gravity of particles
is exactly compensated by their interaction with the dark energy.  
We notice that the $u=0$ homogeneous solution is stable only  for
$0 \leq \xi < 2.04 \ldots$.

We recall that in the vortex case $ \eta=\xi < 0 $ there are
nontrivial solutions for $u(R)$\cite{vortex}.

\bigskip

Let us now study the particle density $\rho(R)$ as a function of
$\xi$. There are three stable cases ($ \kappa_T > 0 $). We illustrate
these cases plotting 
$f(\eta,\xi)$, the surface density in units of $ \frac{N \; T}{V} $ in
fig. \ref{ps}. That is, $f(\eta,\xi)$ vs. $\eta$ for  $\xi = 1,2.04\ldots$,
and $3$.  The densities $\rho(R)$ are plotted as functions of $R$ for
$\xi = 1,2.04\ldots$, and $3$ and chosen values of $\eta$ in
figs. \ref{rho1} to \ref{rho3}.

\begin{itemize}

\item{\bf First case $0 \leq \xi<2.04 \ldots$}

\noindent We illustrate this case choosing the value $\xi=1$. As we
see in fig. \ref{ps} the line $\eta=1$ intercepts $f(\eta,\xi=1)$ just
in one point. This intersection corresponds to the solution
$u(x)=0$. At this point 
$\xi=\eta$ we also have $u_0=0$. We want to notice that when we move 
over the curve $f(\eta,\xi=\mbox{fixed})$ increasing $\eta$, $u_0$ {\bf
always increase}. Therefore, $u_0<0$ above
the point where $u(x)=0$ while  $u_0>0$ below the 
point where $u(x)=0$. This property holds for all values of $\xi$
(for $\xi>1.638 \ldots$  the line $\eta=\xi$ intercepts
$f(\eta,\xi)$ vs. $\eta$ in more than two points, for $1.638
\ldots<\xi<2.04\ldots$ the point $u=0$ is in the upper branch).

\noindent We have for this case $\eta^0(\xi)>\xi$ .
There are two zones of stability in fig. 2 for $f(\eta,\xi=1)$: 

\noindent -the zone between the point $\eta=0$ and the point of
homogeneous density $(\eta=\xi,u=0)$, where  $u_0<0$ and 
$\eta<\xi \; (<\eta^0(\xi))$, which corresponds to a density increasing
with $R$. 

\noindent -the zone between the point of homogeneous density
$(\eta=\xi,u=0)$ and the point of unstability 
$\eta=\eta^0(\xi)$, where  $u_0>0$ and
$\xi<\eta<\eta^0(\xi)$, which corresponds to a  density decreasing with $R$. 

\noindent The gas is unstable in the zone beyond the critical point of
unstability $\eta=\eta^0(\xi)$.  

\noindent We plot in fig. \ref{rho1} the densities $\rho(R)$ vs. $R$ for
$\xi=1$ and $\eta=0.67\ldots, \; \eta=1 $ and $\eta=1.41\ldots $.
The first plot corresponds to a increasing density. 
The dark energy dominates over
gravitational attraction. The second plot corresponds to the uniform
density associated to $u(x)=0$. The density is homogeneous and dark
energy exactly compensates gravitational attraction. The third plot 
corresponds to a decreasing density. The gravitational attraction
dominates over dark energy.  

\item{\bf Second case $\xi=\xi^0=2.04 \ldots$}

\noindent This is the value of $\xi$  where $\eta^0(\xi)=\xi$. This is
a `triple point' since it is defined by two conditions
$$
f(\eta=\xi,\xi)=1 \quad  \mbox{and} \quad \eta^0(\xi)=\xi
$$ 

In this case the point of unstability $\eta=\eta^0(\xi)$ coincides with
the point of homogeneous density $(\eta=\xi^0,u=0)$.
In fig. 2 $f(\eta,\xi=2.04\ldots)$ the gas is stable in the zone
between the point $\eta=0$ and the point of 
homogeneous density $(\eta=\eta^0(\xi^0)=\xi^0,u=0)$,  where  $u_0<0$ and
$\eta<\eta^0(\xi^0)=\xi^0$, which corresponds to an increasing  density. 
The gas is unstable in the zone beyond the critical point of
unstability $\eta=\eta^0(\xi)$.  

\noindent We plot in fig. \ref{rho2p04} the densities $\rho(R)$ vs. $R$ for
$\xi=2.04...$ and $\eta=1$, $\eta=2.04\ldots $.
The first plot exhibits a  density increasing with $R$.
The dark energy dominates over the gravitational attraction. The second
plot corresponds to the point of homogeneous density $u(x)=0$, which
coincides with the point of unstability $\eta=\eta^0(\xi)$. 

\item{\bf Third case $\xi>2.04\ldots$}

\noindent We illustrate this case choosing the value $\xi=3$.
We have $\eta^0(\xi)<\xi$ for this case. The homogeneous solution $u=0$ 
is unstable because $ \kappa_T < 0 $ for it. In the graph
$f(\eta,\xi=3)$ the gas is stable in    
the zone  between the point $\eta=0$ and the point of unstability
$\eta=\eta^0(\xi)$, where  $u_0<0$ and
$\eta<\eta^0(\xi)(<\xi$), which corresponds to an increasing density.

\noindent We plot in fig. \ref{rho3} the densities $\rho(R)$ vs. $R \;$ 
for $\xi=3$ and $ \eta= 1.379\ldots$. 
The plot corresponds to an increasing density. The dark energy dominates
over gravitational attraction.

\end{itemize}

It could be mentioned that the homogeneous solution $u(x)\equiv 0$
that we have for $ \eta = \xi $ is the nonrelativistic analogue of the
static Einstein universe. It was precisely thanks to the introduction
of  the cosmological constant that Einstein found such static and
homogeneous universe.  

\subsection{The Diluted expansion}

We can obtain  power expansions in $\eta$ and $\xi$ for the various
physical quantities from eqs. (\ref{emx2}) and (\ref{dl}). 
These analytic expressions are valid for $\eta \ll 1$ and $\xi \ll
1$. We find,
\begin{equation} \label{eu0}
e^{u_0}=\frac{\eta}{\xi}
\left[1+\frac{3}{10}\;(\eta-\xi)+\frac{51}{1400}\;\xi^2 
-\frac{183}{1400}\; \eta\; \xi+\frac{33}{350}\; \eta^2+O(\eta^3,\eta^2\;
\xi,\eta\; \xi^2,\xi^3) \right] \; . 
\end{equation}
\noindent  Using now eqs.(\ref{f}), (\ref{dl}) and (\ref{eu0}), we obtain
\begin{equation} \label{dvf}
f(\eta,\xi)=1-\frac{\eta-\xi}{5}-\frac{\eta-\xi}{175} \; (\eta+2 \, \xi)
+O(\eta^3,\eta^2 \xi,\eta \;  \xi^2,\xi^3)
\end{equation}

\noindent Eq.(\ref{dvf}) reduces for $\xi=0$ to the
corresponding expression in absence of dark energy (see
\cite{gasn}). For $\eta=\xi$ we retrieve the solution $u=0$ with $f=1$. 
We see comparing with fig.\ref{ps} that the first order approximation
$f(\eta,\xi)=1-\frac{\eta-\xi}{5}$ 
is valid even for {\bf large } $\eta$ and $\xi$ till the vicinity of
the point where the slope becomes vertical. This unexpectedly large
domain of validity may be explained by the smallness ($\frac{1}{175}$)
of the second order coefficient.

We clearly see that the cosmological constant {\bf translates} as a whole
the plot of $f(\eta,\xi=0)$ by an amount  $+\xi$. That is, 
$$
f(\eta,\xi) \simeq f(\eta-\xi,\xi=0)
$$
for $ \eta \lesssim \eta_C(\xi) $.

\bigskip

\section{The limiting case $\xi \gg \eta$}

We consider here the limit where the dark matter dominates over the
ordinary matter ($\xi \gg \eta$). We have a gas of particles where their
self-gravitational interactions are dominated by the interaction
with the dark energy. 
 
The mean field  equation (\ref{Poisson}) becomes linear
$$
 \nabla^2 \Phi({\bf x})=4 \pi \; \xi  \; .
$$
The exact solution for spherical symmetry takes the form,
\begin{equation} \label{fixp}
\Phi(R)=\alpha+\frac{\xi^R}{2}\left(\frac{R}{R_{max}}\right)^2 \; ,
\end{equation}
\noindent where $\alpha$ is independent of $R$.
We recall that $R$ runs in  $0 \leq R \leq R_{max} $ with
$ R_{max}=\left(\frac{3}{4 \pi}\right)^{\frac{1}{3}} $.

The density of matter follows from eqs.(\ref{phi}) and (\ref{fixp}),
\begin{equation} \label{rhoxp}
\rho(R)=e^{\alpha} \; e^{\frac{\xi^R}{2}\left(\frac{R}{R_{max}}\right)^2}
\end{equation}
\noindent Notice that this density increases monotonically with $R$
	  [see sec. IV.C].  

The contrast is given by
$$
C(0,\xi)=\frac{\rho(0)}{\rho(R_{max})}=e^{- \frac{\xi^R}{2}}\;  .
$$
The density normalization (\ref{normalisation}) imposes that
\begin{equation} \label{alfa}
e^{-\alpha(\xi^R)}=3 \; \int_0^1 dy \; y^2  \; e^{\frac{\xi^R}{2} \; y^2} \; .
\end{equation}
This integral can be expressed in terms of the probability integral
$\Phi(x)$ as follows\cite{grad}
$$
e^{-\alpha(\xi)}=\frac{3}{\xi^{\frac32}} \; \left[i \; \sqrt{\frac{\pi}{2}}
\; \Phi\left(i \; \sqrt{\frac{\xi}{2}}\right) + \sqrt{\xi} \;
e^{\frac{\xi}{2}}\right] \; ,
$$
where 
$$
\Phi(x) \equiv \frac{2}{\sqrt{\pi}} \int_0^xdy \; e^{-y^2} \; .
$$

It is instructive to derive these results dropping the particle
self-interaction in the partition function (\ref{part}).
 The Hamiltonian is in this limit,
$$
H = \sum_i \frac{p_i^2}{2 \, m_i^2} - \frac{4\pi \, G \, \Lambda}{3}
\,\sum_i m_i \; q_i^2  
$$
The coordinate partition function $ Z_{coor} $ then {\bf factorizes}
and we find from eq.(\ref{part}),
$$
Z=\frac{V^{N}}{N!} \left(\frac {m T}{2 \pi}\right)^{3 N/2} \; \; e^{-N
\alpha(\xi^R)} \; . 
$$
where $ \alpha(\xi^R) $ is given by eq.(\ref{alfa}) after using the
dimensionless coordinates $ {\vec r} \equiv {\vec q}/L $.

\noindent The density at the boundary (\ref{densper}) takes here the form
$$
f(0,\xi^R)=\rho(R_{max})=e^{\alpha(\xi^R)} \; e^{\frac{\xi^R}{2}}
$$
We plot in fig. \ref{fpure} the density at the boundary versus $\xi$.
We give below the thermodynamic quantities in terms of $f(0,\xi^R)$ and
$\xi^R$. That is, free energy, potential energy, pressure
at the boundary, entropy, specific heat at constant volume and
isothermal compressibility of the gas:   

\begin{figure}[htbp]
\rotatebox{-90}{\epsfig{file=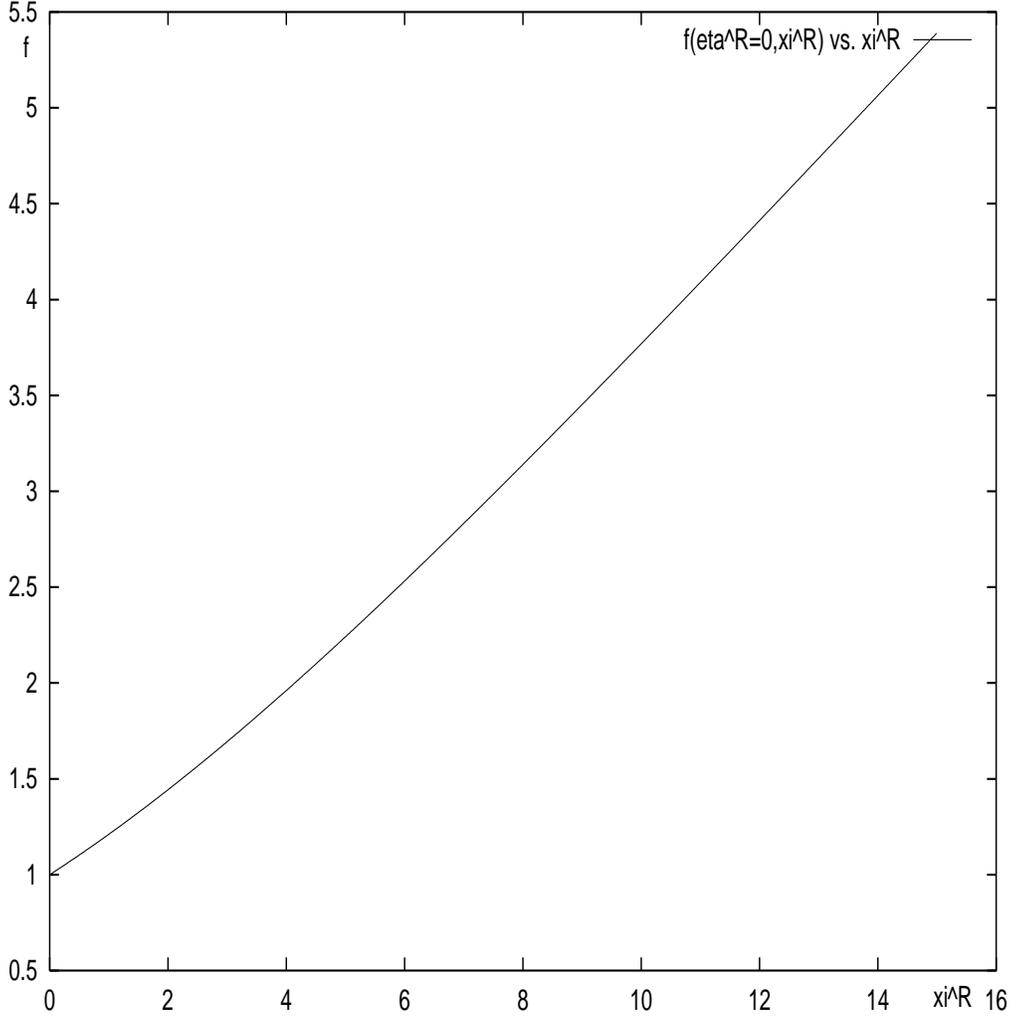,width=14cm,height=14cm}}
\caption{The density at the boundary $ f(\eta^R=0,\xi^R) $ versus $\xi^R$
when the effects of the dark matter dominate  over the selfgravitation.} 
\label{fpure}
\end{figure}

\begin{eqnarray} \label{thxp}
&&\frac{F-F_0}{NT}=\alpha(\xi^R)=\ln{f(0,\xi^R)}-\frac{\xi^R}{2} \quad , \quad
\frac{E_P}{NT}=\frac{3}{2}\left[1-f(0,\xi^R)\right] \cr \cr 
&&\frac{PV}{NT}=f(0,\xi^R) \quad , \quad
\frac{S-S_0}{N}=\frac{3}{2}[1-f(0,\xi^R)]-\ln{f(0,\xi^R)}+\frac{\xi^R}{2} \; 
\cr \cr 
&& c_v=\frac{3}{2} + \frac{3}{4} \; \xi^{R} \;
f(0,\xi^R)+\frac{3}{2}[1-f(0,\xi^R)][1+\frac{3}{2}f(0,\xi^R)]  \; \cr
\cr  
&& \frac{1}{\kappa_T} = f(0,\xi^R) \; \left[f(0,\xi^R)-\frac{\xi^R}{3}\right]
\end{eqnarray}
$F_0$ and $S_0$ are respectively the free energy and the entropy of the
perfect gas. The second and third relations in eqs.(\ref{thxp}) yield
virial theorem,
$$
\frac{PV}{NT}=1-\frac{2}{3} \; E_P \; .
$$
which is the limiting case of eq.(\ref{virial}) for $ \eta = 0 $.

We checked that $c_v$ and $\kappa_T$ are positive for all
$\xi^R\geq 0$. The gaseous phase is thus stable in this limit. 

\bigskip

\noindent We compute the asymptotic behavior of the thermodynamic
functions for $\xi^R \to \infty$
\bea
&&\frac{PV}{NT}=f(0,\xi^R)=\frac{\xi^{R}}{3}+\frac{1}{3}+\frac{2}{3 \; \xi^R}
+\frac{4}{{\xi^{R}}^2} +O\left(\frac{1}{{\xi^R}^3}\right) \cr\cr && 
\frac{F-F_0}{NT}=-\frac{\xi^{R}}{2}+\ln{\frac{\xi^{R}}{3}}+\frac{1}{\xi^R}
+\frac{3}{2 {\xi^R}^2}+O\left(\frac{1}{{\xi^R}^3}\right)\cr\cr
&&\frac{E_P}{NT}=-\frac{\xi^{R}}{2}+1-\frac{1}{\xi^R}-\frac{6}{{\xi^R}^2}
+O\left(\frac{1}{{\xi^R}^3}\right)\quad , \quad
\frac{S-S_0}{N}=-\ln{\frac{\xi^{R}}{3}}+1-\frac{2}{\xi^R}-\frac{15}{2
  {\xi^R}^2}+O\left( \frac{1}{{\xi^R}^3}\right)\cr\cr
&&c_v=\frac{5}{2}-\frac{7}{2 \xi^R}+O\left(\frac{1}{{\xi^R}^2}\right)
\quad , \quad
\kappa_T=\frac{9}{\xi^R}-\frac{27}{{\xi^R}^2}-\frac{63}{{\xi^R}^3}
+O\left(\frac{1}{{\xi^R}^4}\right)
\eea
and for $\xi^{R} \to 0$
\bea
&&\frac{PV}{NT}=f(0,\xi^R)=
1+\frac{\xi^{R}}{5} +\frac{2 {\xi^{R}}^2}{175}+O({\xi^{R}}^3)
\quad , \quad
\frac{F-F_0}{NT}= -\frac{3 \xi^{R}}{10}-\frac{3 {\xi^{R}}^2}{350}
  +O({\xi^{R}}^3)\cr\cr 
&&\frac{E_P}{NT}=-\frac{3 \xi^{R}}{10}-\frac{3 {\xi^{R}}^2}{175}+O({\xi^{R}}^3)
\quad , \quad
\frac{S-S_0}{N}=-\frac{3 {\xi^{R}}^2}{350}+O({\xi^{R}}^3)\cr\cr
&& c_v=\frac{3}{2} +\frac{3 \; {\xi^{R}}^2}{175}+O({\xi^{R}}^3)\quad , \quad
\kappa_T=1-\frac{\xi^R}{15}+\frac{13}{1575} {\xi^R}^2+O({\xi^{R}}^3)
\eea
The particle density for  $\xi^R \to \infty$ takes the form,
$$
\rho(R)= \frac{\xi^R + 1}{3} \; e^{- \frac{\xi^R}{2}\left[1
    -\left(\frac{R}{R_{max}}\right)^2 \right]}  \; .
$$
Notice that the $\xi^R \to \infty$ corresponds to keeping $\Lambda$
(and $\eta$) fixed in the thermodynamic limit [see eq.(\ref{defxi})].

\section{Discussion and Conclusions}

The behaviour of the self-gravitating gas is significantly influenced 
by the cosmological constant (or not) depending on the value of the ratio
$$
R_{\Lambda} \equiv \frac{\xi}{\eta} = \frac{2 \, \Lambda \; V}{m \; N} = 2 \;
\left(\frac{\mbox{dark \; energy}}{\mbox{mass}}\right)_V = 2 \;
\frac{\Lambda}{\rho_V}   
$$
where $\rho_V \equiv N \; m /V $ is the average mass density in the
volume $V$ and where the subscript $\left( \right)_V$ indicates mass and dark
energy inside the volume $V$.

We display in Table 2 the typical value of $R_{\Lambda}$ for some relevant
astrophysical objects. Except for the universe as a whole $R_{\Lambda}$ takes
small values. For such values of $R_{\Lambda}$ the effect of the cosmological
constant is negligible as we see from the previous sections. 
The smallness of  $R_{\Lambda}$ stems from the tiny value of $\Lambda
= 0.663 \; 10^{-29}$g/cm$^3$\cite{lam}. The particle mass density is
much larger than $\Lambda$ in all situations except for the universe as a
whole. However, the nonrelativistic and equilibrium treatment does not
apply for the universe as a whole.

\bigskip

\begin{tabular}{|l|l|}\hline
  & \\ 
Object   & $R_{\Lambda}$ \\ 
   & \\ \hline
 Stars    & $\sim 10^{-29}$\\ \hline 
 Supergiants & $\sim 10^{-22}$\\ \hline 
 Galaxies  & $\sim  10^{-6} -10^{-5}$ \\ \hline 
Clusters of Galaxies & $\sim  10^{-2}$ \\ \hline 
The Universe & $\sim  4$ \\ \hline
\end{tabular}

\bigskip

{TABLE 2. The ratio $R_{\Lambda} = \frac{\xi}{\eta}$ for some
  astrophysical objects}

\bigskip

For the cold clouds in the interstellar medium (ISM) considered in
refs.\cite{gasn,gas2,gas3} it is more convenient
to express $ \xi $ as
$$
\xi = 1.2 \, 10^{-6} \; \, \frac{m(A)}{T(K)} \; \left[ L(pc) \right]^2
$$
where the mass of the particles $m$ is expressed in units of the
hydrogen mass and $T$ in Kelvins. For typical  cold clouds in the ISM
we have $ 1 < L(pc) < 100 $ and $ T \sim 5 $ \cite{gas2}. This yields
$$
\xi \sim 10^{-3} \quad \mbox{to} \quad 10^{-7} \; .
$$
while $ \eta \lesssim 1 $. Again, we find $R_{\Lambda} \ll 1 $. 

\bigskip

The isothermal sphere in the presence of a cosmological constant has
been considered in refs.\cite{tis}. The authors in ref.\cite{tis}
restrict themselves to the case where the gravitational force points
inward  in all points. This happens when $ \eta  >  \xi $ (using our
notation). As showed here, when the dark energy dominates, there are
stable solutions for $  \xi >  \eta $ where  $\rho(R)$ {\bf increases}
with $R$. In these situations where the dark energy dominates the
gravitational force points outward in all points. 

It must be noticed that the mean field equation (\ref{reducedeq})
admits solutions for which the density is not a monotonous function of
$R$. However, such solutions are unstable since they correspond to negative
values for the compressibility $ \kappa_T$.

The effect produced by the cosmological constant is  of an opposite
nature to the Jeans' or gravothermal instabilities where the
self-gravitating gas collapses due to the mutual attraction. Namely,
the points C and MC in figs. 1 and 2.

\section{Appendix}

\subsection{Appendix A}

\noindent
The goal of this appendix is to compute in the spherically symmetric
case the following integrals 
$$
A=
\int {\rm d}^3{\vec x} \;e^{\Phi({\vec x})}\; \left[\Phi({\vec x})\;
-\frac{2 \; \pi}{3} \; \xi \; x^2 \; \right] \quad  , \quad
B=-\int {\rm d}^3{\vec x} \;e^{\Phi({\vec x})}\; \left[\Phi({\vec x})\;
+\frac{\pi}{3} \; \xi \; x^2 \; \right] \; .
$$ 
Using eq.(\ref{ecsd}), we obtain 
\begin{eqnarray} \label{Ap}
A&=&-\frac{1}{4 \pi \eta} \int {\rm d}^3{\vec x} \; \Phi({\vec x})\;
\;\nabla^2 \Phi({\vec x}) 
+\frac{1}{6} \; \frac{\xi}{\eta} \; \int {\rm d}^3{\vec x} \; x^2
\;\nabla^2 \Phi({\vec x})  \nonumber \\
&&+\frac{\xi}{\eta} \int {\rm d}^3{\vec x} \; \Phi({\vec x})-\frac{2
  \pi}{3}\frac{\xi}{\eta} \; \xi  \int {\rm d}^3{\vec x} \;x^2
\nonumber \\ 
B&=&\frac{1}{4 \pi \eta} \int {\rm d}^3{\vec x} \; \Phi({\vec x})\;
\;\nabla^2 \Phi({\vec x}) 
+\frac{1}{12} \; \frac{\xi}{\eta} \; \int {\rm d}^3{\vec x} \; x^2
\;\nabla^2 \Phi({\vec x}) 
 \nonumber \\
&&-\frac{\xi}{\eta} \int {\rm d}^3{\vec x} \; \Phi({\vec x})-\frac{
\pi}{3}\frac{\xi}{\eta} \; \xi  \int {\rm d}^3{\vec x} \;x^2 \nonumber 
\end{eqnarray}
The terms in the first lines of $A$ and $B$ can be integrated twice by parts 
with the following result in the spherically symmetric case,
\begin{eqnarray} \label{ApBp} 
A&=&\left(1-2 \frac{\xi}{\eta}\right) \ln{f(\eta,\xi)} -\frac{\xi^R}{2}\left( 1 -
\frac{2 \, \xi^R}{5 \, \eta^R} \right) +I+2 J  \nonumber\\
B&=&-\left(1- \frac{\xi}{2 \; \eta} \right) \; \ln{f(\eta,\xi)}-\frac{\xi^R}{4}\left( 1 -
\frac{2 \, \xi^R}{5 \, \eta^R} \right) -I -\frac{J}{2}\; ,
\end{eqnarray}
where $f(\eta,\xi)$ is defined by eq.(\ref{densper}) and 
\begin{equation} \label{IJ}
I \equiv \frac{1}{\eta} \int_{0}^{R_{max}} {\rm d}R \; R^2 \;
[\Phi'(R)]^2 \quad  , \quad
J \equiv \frac{4 \, \pi \, \xi}{\eta} \int_{0}^{R_{max}}{\rm d}R \;R^2
\;\Phi(R)  \; .
\end{equation} 
We compute now $I$ in terms of $J, \; f, \; \eta$ and $\xi$.
Let us consider the function 
$$ 
B(R)\equiv R^3  \; e^{\phi(R)} \; .
$$
We take the derivative with respect to $R$ and use  eq.(\ref{Poissonrad}),
$$
B'(R)=3 \; R^2 \; e^{\Phi(R)}-\frac{1}{4 \pi \eta} \; R^3 \;\Phi'(R)
\; \Phi''(R)-\frac{1}{2 \pi \eta} \; R^2 \; [\Phi'(R)]^2+R^3 \;
\frac{\xi}{\eta} \; \Phi'(R) \; .
$$
We now integrate $B'(R)$ between $0$ and $R_{max}$.
Integrating by parts the second and the fourth terms and using
eqs.(\ref{vincu2}) and (\ref{IJ}) yields,
\begin{equation} \label{I}
I=-6 \;  J+6\; [1-f(\eta,\xi)]+6\; \frac{\xi}{\eta} \; \ln{f(\eta,\xi)}-\eta^R+
\xi^R \left( 2 -\frac{\xi^R }{\eta^R } \right) \; .
\end{equation} 
We finally get for $A$ and $B$ using eqs.(\ref{ApBp}) and  (\ref{I})
\begin{eqnarray} \label{AsBs}
A&=&6[1-f(\eta,\xi)]+\left(1+4 \; \frac{\xi}{\eta}\right)
\ln{f(\eta,\xi)}-\eta^R+ 
\xi^R \left(\frac{3}{2} - \frac{4 \xi}{5\eta}\right) -4 J
\nonumber\\ 
B&=&6[f(\eta,\xi)-1]-\left(1+ \; \frac{11 \; \xi}{2 \;
  \eta}\right)\ln{f(\eta,\xi)} 
+\eta^R-\xi^R \left(\frac{9}{4}-\frac{11 \; \xi}{10 \; \eta}\right)
+\frac{11}{2} \; J \; . 
\end{eqnarray}

\subsection{Appendix B}

\noindent
In this appendix, we will compute the specific heat at constant volume.
Using eqs.(\ref{etaxi}), (\ref{entr}) and (\ref{defcv}), we find
\bea 
&&c_v=6 \; T \left( \frac{\partial f}{\partial T}
\right)_V-\frac{1}{\eta^R} \; \left( \eta^R - \xi^R\right)^2
-\left(1+5 \; \frac{\xi^R}{\eta^R} \right) \frac{T}{f} \; \left(
\frac{\partial f}{\partial T} \right)_V  +\cr \cr
&&+20 \pi \; \frac{\xi^R}{\eta^R} \; \left.T \frac{\partial}{\partial
  T}\right|_V  \int_0^{R_{max}}{\rm d}R \; R^2 \; \Phi(R)
\eea
\noindent
Using eqs.(\ref{f}) we calculate the partial derivative of $f$ with
respect to $T$,
$$
T \left( \frac{\partial f}{\partial T} \right)_V=T \left(
\frac{\partial u}{\partial T} \right)_V(x=\sqrt{3 \; \xi^R},u_0) \;
f. 
$$
\noindent
We obtain from eqs.(\ref{etaxi}), (\ref{emx2}) and (\ref{g}),
\begin{equation} \label{Tdf}
T \left( \frac{\partial f}{\partial T} \right)_V=\left[ \frac{1}{2}
(\eta^R-\xi^R)+ 
T \left( \frac{\partial u_0}{\partial T} \right)_V \; g(x=\sqrt{3 \;
  \xi^R},u_0) \right] f \; . 
\end{equation}

\noindent
Using eqs.(\ref{etaxi}), (\ref{transf}) and (\ref{g}) we find
\bea
&&\left.T \frac{\partial}{\partial T}\right|_V
\int_0^{R_{max}}{\rm d}R \; R^2 \; \Phi(R) =-\frac{3}{8 \; \pi}
\frac{1}{(3 \; \xi^R)^{3/2}} 
\int_0^{\sqrt{3 \; \xi^R}}{\rm d}x \; x^3 \; \left( \frac{ \partial
  u}{ \partial x} \right)(x,u_0) \cr  \cr 
&&+\frac{3}{4 \; \pi} \;
T \left( \frac{\partial u_0}{\partial T} \right)_V \; 
 \frac{1}{(3 \; \xi^R)^{3/2}}
\int_0^{\sqrt{3 \; \xi^R}}{\rm d}x \; x^2 \; g(x,u_0) \; .
\eea
\noindent
Integrating by parts the first integral  of the right member of the
last equation, we obtain 
\begin{eqnarray} \label{Tdi}
&&\left.T \frac{\partial}{\partial T}\right|_V
\int_0^{R_{max}}{\rm d}R \; R^2 \; \Phi(R)=    
\frac{9}{8 \; \pi} \frac{1}{(3 \; \xi^R)^{3/2}}
\int_0^{\sqrt{3 \; \xi^R}}{\rm d}x \; x^2 \; u(x,u_0) \nonumber\\
&&-\frac{3}{8 \; \pi} \; u(\sqrt{3 \; \xi^R},u_0)
+\frac{3}{4 \; \pi} \;
T \left( \frac{\partial u_0}{\partial T} \right)_V \; 
 \frac{1}{(3 \; \xi^R)^{3/2}}
\int_0^{\sqrt{3 \; \xi^R}}{\rm d}x \; x^2 \; g(x,u_0)
\end{eqnarray}
\noindent
Using eqs.(\ref{transf}), (\ref{Tdf}) and (\ref{Tdi}) we express $c_v$ as
in eq.(\ref{cv2})

\section{Acknowledgments}

We thank N. S\'anchez for useful discussions.

\end{document}